\documentclass[aps,pre,twocolumn,showpacs,10pt]{revtex4-1}
\usepackage{color}
\usepackage{amsmath}
\usepackage{amsfonts}
\usepackage{amssymb}
\usepackage{graphicx}

\begin{document}
\title{Connectivity, Dynamics, and Structure in a Tetrahedral Network Liquid}
\author{S{\'a}ndalo Rold{\'a}n-Vargas$^{1,2}$}
\email{sandalo@pks.mpg.de} 
\author{Lorenzo Rovigatti$^{3,4}$}
\author{Francesco Sciortino$^2$}
\affiliation{$^1$Max Planck Institute for the Physics of Complex Systems,  D-01307, Dresden, Germany,\\ 
$^2$Department of Physics, Sapienza, Universit\`{a} di Roma, Piazzale Aldo Moro 2, I-00185, Roma, Italy,\\
$^3$Faculty of Physics, University of Vienna, Boltzmanngasse 5, A-1090 Vienna, Austria\\
$^4$Rudolf Peierls Centre for Theoretical Physics, 1 Keble Road, Oxford, OX1 3NP, UK}

\begin{abstract}
We report a detailed computational study by Brownian Dynamics simulations of the structure and dynamics of a liquid of patchy particles which develops an amorphous tetrahedral network upon decreasing temperature. The highly directional particle interactions allows us to investigate the system connectivity by discriminating the total set of particles into different populations according to a penta-modal distribution of bonds per particle. With this methodology we show how the particle bonding process is not randomly independent but it manifests clear bond correlations at low temperatures. We further explore the dynamics of the system in real space and establish a clear relation between particle mobility and particle connectivity. In particular, we provide evidence of anomalous diffusion at low temperatures and reveal how the dynamics is affected by the short-time hopping motion of the weakly bounded particles. Finally we widely investigate the dynamics and structure of the system in Fourier space and identify two quantitatively similar length scales, one dynamic and the other one static, which increase upon cooling the system and reach distances of the order of few particle diameters. We summarize our findings in a qualitative picture where the low temperature regime of the viscoelastic liquid is understood in terms of an evolving network of long time metastable cooperative domains of particles.         
\end{abstract}

\maketitle

\section{INTRODUCTION}

Most of the distinctive features of the equilibrium relaxation in disordered systems support the existence of microscopic cooperative motion~\cite{adam_gibbs,binder_walter,biroli_berthier}.  Examples are the spatially heterogeneous dynamics~\cite{walter_heterogeneities,berthier_heterogeneities,berthier_science,ediger} (where large regions of the system relax significantly faster, or slower, than the average), the non-Gaussian distributions of displacements~\cite{pinaki_non_gaussian,nat_mat_non_gaussian_granick,hs_non_gaussian} (resulting from averaging coupled and non-equally distributed individual displacements), or the non-Debye decays of the time correlation functions~\cite{angell_relaxation,ediger} (which again represent a manifestation of the system heterogeneous relaxation). With these empirical features we have composed a general \textit{dynamic} catalog to characterize the slow relaxation of viscous liquids. Nonetheless, the problem of whether or not this dynamic phenomenology relies on an underlying structure still remains open ~\cite{binder_walter,biroli_berthier,biroli_garrahan,debenedetti_Stillinger,lubchenko_wolynes}.\\

So far this phenomenology has been found in a large variety of systems such as molecular liquids~\cite{hansen,phillips}, colloidal systems~\cite{pusey_megen,pham_science,pham_pre}, polymer fluids~\cite{matsuoka_polymer,phillies_stretched}, granular media~\cite{diffusion_granular,jaeger_granular}, or spin glasses~\cite{mezard_parisi_virasoro}. In general these systems have been categorized by the  wide notions of \textit{glass} and \textit{gel} depending on their particular density and temperature behaviors as well as on their specific microscopic properties~\cite{binder_walter,biroli_berthier,biroli_garrahan,parisi_zamponi,debenedetti_Stillinger,cipelletti_2005,zaccarelli_gels}. Thus, we discriminate between strong and fragile glasses according to their viscosity temperature dependence (Arrhenius or super-Arrhenius)~\cite{angell,bohmer_angell} or between repulsive and attractive glasses~\cite{pham_science,pham_pre} depending on the character of the microscopic interactions present in the system. Often in a non-rigorous  manner, we also distinguish glasses from gels according to macroscopic criteria such as viscoelasticity or solid-like behavior at low densities (gels) as well as microscopic criteria such as the existence of an amorphous  network structure, which is typical of gelling systems~\cite{zaccarelli_gels,cipelletti_2005}. Apart from these available classifications, some of them phenomenological, microscopic canonical models can help us to not only develop novel applications but to better understand the microscopic mechanisms involved in the cooperative dynamics, and in its hypothetical structural origin, present in gels and glass forming liquids~\cite{binder_walter,biroli_berthier,zaccarelli_gels,cipelletti_2005}.\\
      
In the present work we study one of these models. By means of Brownian Dynamics (BD) simulations, we investigate the equilibrium dynamic and static properties of a gelling system which develops an amorphous tetrahedral network upon decreasing temperature. In particular, we report a comprehensive study on the interplay between connectivity, dynamics, and structure which results in a qualitative microscopic picture for the viscoelastic nature of the low temperature liquid.\\ 

The system is constituted by tetravalent \textit{patchy particles}, that is, particles with sticky spots on their surface which provide a strongly directional interaction with fixed valence~\cite{dnapatchy1,granickpatchy2011,glad1,flavioNC}. Far from being merely a theoretical or computational idealization, patchy particles are nowadays amenable to experimentation~\cite{dnapatchy1,granickpatchy2011,achillebook,granickpatchy2012}. Our particular realization consists of particles with four attractive patches tetrahedrally distributed on the particle surface~\cite{John_valence,rovigatti_molphys}. Despite its simplicity, this and similar models~\cite{KF,granick,romanosoftmatter,achillebook,widepatches} have already shown the capability for capturing some of the fundamental structural features of different classical systems with amorphous tetrahedral structure such as atomistic models of water, silicon, or silica~\cite{ivan_patchy_water,water1_ff,water2_ff,binder_walter}. In general, due to their highly versatile functionality, these models have not only the potential for promising  applications~\cite{vitrimers_ff,gel_inverso,phase_diagram_inverso} but they can also allow us to reach a deeper understanding of some of the intriguing phenomena manifested in gel and glassy systems ~\cite{zaccarelli_gels,equilibriumgels1}.\\

Here we take advantage of the highly directional interactions to analyze the system cooperativity by describing the potential energy in terms of a distribution of bonds per particle. We widely exploit this distribution of bonds to partition the system into particle populations. This discrimination allows us to establish a clear relation between connectivity and particle mobility. Indeed, we study the dynamics of these particle populations in real and Fourier space by covering several temperatures within a large range of spatial scales and times, from the pure diffusive regime to the heterogeneous low temperature dynamics. Of particular interest are the results concerning the self and collective dynamics of the system in Fourier space. From these results we show the decoupling between self and collective dynamics at low temperatures within a certain range of the wavevector $q$. At even smaller values of $q$ we identify a \textit{dynamic} length scale which increases upon cooling the system and signals the emergence of cooperative domains  of particles. This dynamic length scale reaches distances of the order of few particles diameters. Thanks to the study of the structure factor of the different particle populations, we are also able to demonstrate the existence of a second long-range length scale of \textit{static} nature which is quantitatively similar to the previous dynamic length scale. This result suggests a clear connection between structure and dynamics at large spatial scales. We synthesize all these findings in a microscopic picture where the viscoelastic nature of the low temperature liquid would be the result of a viscous flow of cooperative domains of particles coupled to the network elasticity, which is in its turn mediated by the inter-domain connections.\\

The rest of the paper is organized as follows. In section II we present the model. Section III contains the results and consists of three parts: In the first part we present those results concerning the system connectivity. In the second part we explore the dynamics of the system in real and Fourier space. The third part is devoted to the system structure. Finally in Section IV we summarize our main results and present our conclusions. We also include an Appendix which contains technical results and information concerning the structure and dynamics of the system in Fourier space.\\
 
\section{MODEL}

We perform three dimensional BD simulations of tetravalent patchy particles in the canonical ensemble. We fix the number of particles $N=10000$ and the simulation volume $V=L^3$, where $L=25.98\;\sigma$, being $\sigma$ the particle hard sphere-like diameter. With these choices the number density is $\rho=N/V=0.57 \sigma^{-3}$. For this value of the density, our system is included in the so-called optimal network density region, \textit{i.e.} the region at which an unstrained fully bounded network can form at low temperature ~\cite{fully_bonded_tetra}. Under these conditions the dynamics of the system exhibits an Arrhenius behaviour, which is the defining characteristic of strong glass formers. As a result, differently from fragile glass formers, the quantities we evaluate do not show any sign of divergence at finite temperature.\\

\begin{figure}[tb]
\center
\includegraphics[width=0.9\linewidth]{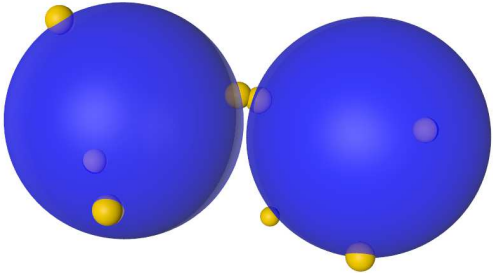} 
\caption{Sketch of two particles with four patches tetrahedrally distributed on their surface. The intersection between two patches, small yellow spheres, results in an attractive interaction given by the interaction potential $V_P(1,2)$ (see Equation (3)).} 
\center
\label{Fig_cartoon} 
\end{figure}

The interaction potential we use comprises a spherical steep repulsion and a short-range attraction that depends on the distance and on the relative orientation between each pair of patches decorating the particles (See Fig.~\ref{Fig_cartoon}). More precisely, the interaction between a generic pair of particles $1$ and $2$ is given by

\begin{equation}
V(1,2) = V_{CM}(1,2) + V_P(1,2)
\end{equation}
\\
where $V_{CM}(1,2)$ is the repulsive interaction between particles $1$ and $2$  whereas $V_P(1,2)$ represents the attractive interaction between the patches of particles $1$ and $2$. Both interactions are modeled as follows:

\begin{equation} 
V_{CM}(12) = {\left( \frac{\sigma}{r_{12}}\right)}^{m} \,\,\,\,\,\,\,\,\
\end{equation}

\begin{equation}
V_{P}(12) = -\sum_{i=1}^{M}\sum_{j=1}^{M}\epsilon  
\exp \left[ -\frac{1}{2}{\left( \frac{r_{12}^{ij}}{\alpha} \right)}^{n} \right] \,\,\,\,\,\,\,\,\,\,\,\,\
\end{equation}
\\

\noindent
Here $r_{12}$ is the distance between the centers of mass of particles $1$ and $2$, $r_{12}^{ij}$ is the distance between patch $i$ on particle $1$ and patch $j$ on particle $2$, and $M$ is the number of patches per particle, which here we take as $M =4$, being the four patches tetrahedrally distributed on the surface of the particles. Exponents in $V_{CM}(1,2)$ and $V_P(1,2)$ are taken as $m=200$ and $n=10$ to resemble the functional behaviors of a hard sphere and a square well interaction respectively. We select $\alpha=0.12 \sigma$ as the patch linear size to ensure no more than one bond per patch, whereas $\epsilon=1.001$ is chosen in such a way that the minimum of the attractive part of the potential energy in a bounded configuration is $u_0 \equiv \min {V_{P}(12)}=-1$. Temperature, $T$, is measured in units of the potential well (where Boltzmann's constant, $k_B$, is taken as $1$) whereas the time unit is $\sigma\sqrt{mu_0}$, being $m=1$ the mass of the particles. To integrate the equations of motion we use a Velocity Verlet algorithm with a fixed time step $\delta t=0.001$ (See for technical details Ref.~\cite{John_valence}). All simulations were performed with the oxDNA simulation package running on GPUs~\cite{rovigatti_gpu_comparison}.\\

The potential employed in this work has been used in the past as a model for patchy particles ~\cite{John_valence,rovigatti_molphys}. Here, for the first time, and thanks to the increased computer power available, we are able to gain a much deeper insight on the dynamical behavior of the system by looking at the individual contributions of populations of particles. In addition, we have investigated several temperatures within the range $T \in [0.1025, 0.25]$, covering a slowing down of the dynamics of more than four orders of magnitude. 

\section{RESULTS}

\subsection{Connectivity}
\vspace*{0.2cm}
\begin{figure}[tb]
\center
\includegraphics[width=0.9\linewidth]{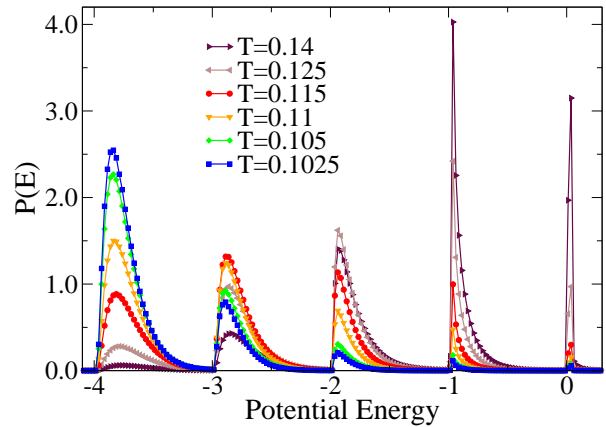} 
\caption{Probability density, $P(E)$, of finding a particle with potential energy $E$ at different temperatures. We clearly observe five main peaks that can be interpreted in terms of the number of bonds per particle. Note that there are some particles with positive potential energy since the repulsive part of the potential energy is not a pure hard-sphere interaction.} 
\center
\label{Fig_Probability_Energy} 
\end{figure}

Despite its continuous potential energy, the system is characterized by construction by a highly directional short-range attraction very similar to that present in square-well-like models whose dynamic and thermodynamic behaviors have been previously investigated~\cite{KF,granick,romanosoftmatter,achillebook,widepatches,phase_diagram_inverso}. In addition, the continuous potential we use allows us to investigate the dynamic behavior at low temperatures.  The highly directional interaction imposed by $V_P(1,2)$, Eq. (3), through a short patch linear size, $\alpha$, induces a probability distribution, $P(E)$, of finding a particle with potential energy $E$ which can be directly understood in terms of a probability distribution of bonds. In this respect, Fig.~\ref{Fig_Probability_Energy}, which synthesizes the $T$-evolution of the system connectivity, shows $P(E)$ at different temperatures. The distribution is characterized by five well-resolved and non-overlapping peaks that can be discretized as a penta-modal distribution. We clearly see how, upon cooling the system, there is a progressive decreasing of the potential energy per particle. In particular, the peak at positive energies corresponds to the population of unbounded particles which from now on we will refer as {\it monomers}.\\

\begin{figure}[tb]
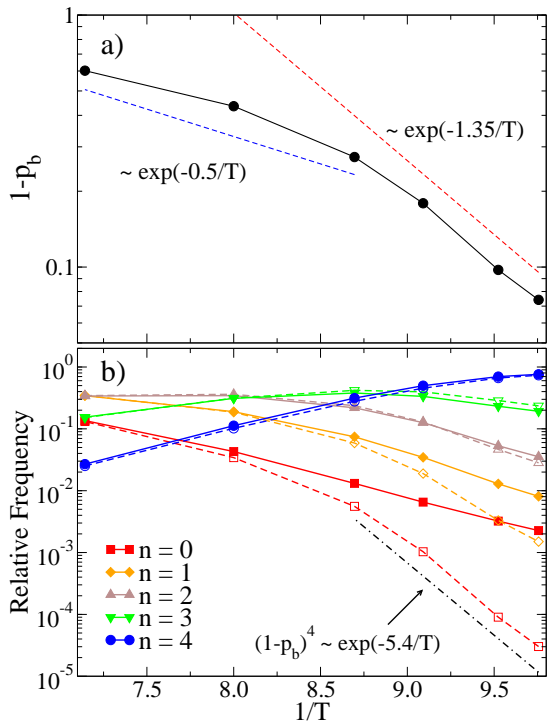

\center
\includegraphics[clip,width=0.4\textwidth]{Unbounded_Patch.eps}
\includegraphics[clip,width=0.4\textwidth]{Bond_T.eps}\\
\caption{a) Probability of observing an unbounded patch, $1-p_b$, as a function of $1/T$ (black line with solid circles). The figure also includes two Arrhenius laws with activation energies $1.35$ (red dashed line) and $0.5$ (blue dashed line). b) Relative frequency of particles with $n$ bonds, ${\cal N}_n$, ($n\in{\lbrace0,1,2,3,4\rbrace}$) as a function of $1/T$. Solid lines with full symbols represent simulation data whereas dashed lines with empty symbols stand for the theoretical prediction, ${\cal N}_n^{Binomial}$, as obtained from an independent bond formation process according to Equation (5). In particular, the figure highlights the theoretical prediction for the expected frequency of monomers, ${\cal N}_0^{Binomial} = (1-p_b)^4$, (dash-dotted line), which follows an Arrhenius law with an activation energy of $4 \times 1.35 = 5.4$ (See a)). }
\center
\label{Fig_Connectivity} 
\end{figure}

To transform $P(E)$ into a discrete penta-modal probability distribution of bonds per particle we merely normalize the area below each peak by the total area of the distribution to obtain an accurate estimation of the fraction (or relative frequency) of particles, ${\cal N}_n$, with a given number of bonds, $n$ ($\in{\lbrace0,1,2,3,4\rbrace}$). Thus, ${\cal N}_n$ is formally determined by:\\

\begin{equation}
{\cal N}_n = \left\{
  \begin{array}{lr}
    \frac{\int_{-n}^{-n+1} P(E) \mathrm{d} E}{\int_{-4}^{\infty} P(E) \mathrm{d} E}     :  n > 0 \\
    \\
    \frac{\int_{0}^{\infty} P(E) \mathrm{d} E}{\int_{-4}^{\infty} P(E) \mathrm{d} E}      :   n = 0 
  \end{array}
\right.
\end{equation}
\\

From the resulting discrete distribution of bonds we can immediately estimate the mean number of bonds per particle, ${\cal N}_b$, as ${\cal N}_b \equiv \sum_{n=0}^4 n {\cal N}_n$, as well as the corresponding bond probability per patch, $p_b \equiv {\cal N}_b/4$. In this respect, Fig.~\ref{Fig_Connectivity}a)  shows the $T$-dependence of the fraction of unbounded patches $1-p_b$.  At low $T$, $1-p_b$ follows an apparent Arrhenius law with an activation energy of about $1.35$. We notice that close to the ground state ($1-p_b \approx 0$) theoretical predictions based on Wertheim theory (in particular on the law of mass action between bounded and unbounded pairs) suggest an Arrhenius law but with an activation energy of $0.5$, that is, half of the depth of the potential well~\cite{Nezbeda_1,Nezbeda_2,fully_bonded_tetra}. This large deviation between simulation and theory, already observed in patchy models with highly directional interaction~\cite{fully_bonded_tetra}, suggests a significant breakdown of the main theoretical assumption which considers an independent (random) bond formation process. Although this deviation still demands a theoretical understanding, it points out that connectivity in the model (in particular at low $T$) is clearly influenced by a correlation between bonds. To provide further evidence of the deviation between an independent bonding process approach and data coming from the simulation, we can consider a simple Binomial distribution of bonds per particle, ${\cal N}_n^{Binomial}$, which, by definition, is based on the independence between the different bonds of a given particle~\cite{florybook}:

\begin{equation}
{\cal N}_n^{Binomial}=\frac{4!}{n!(4-n)!}p_b^n(1-p_b)^{4-n}  
\end{equation}
\\

\noindent  
Here, in order to evaluate ${\cal N}_n^{Binomial}$ at any $T$,  we take $p_b$ as obtained from our simulation. To  precisely quantify the difference between the observed distribution of bonds per particle with that expected from an independent bond formation process, we compare in  Fig.~\ref{Fig_Connectivity}b) ${\cal N}_n$ for the whole set of populations ($n \in{\lbrace0,1,2,3,4\rbrace}$) with ${\cal N}_n^{Binomial}$ within the explored $T$ range. Theory and simulation data only coincide at high $T$, typically above the percolation temperature $T\cong0.12$ ~\cite{bianchi_percolation}, when there is about one bounded patch per particle (${\cal N}_b \approx 1$). Even though each distribution follows its own Arrhenius behavior at low $T$, the data show in a very clear way how the binomial approximation significantly worsens upon cooling the system, especially in regard to the population of monomers, ${\cal N}_0$. In fact, the fraction of monomers at low $T$ is significantly larger than that expected for an independent (binomial) bond formation process. This comparison indicates a separation of the population of unbounded particles from the population of fully bounded particles in the simulation which is larger than that expected for an independent bond formation process.\\ 

The large statistical deviation between theory and simulation data can be interpreted in terms of a spatial localization of the unbounded particles, where the presence of an unbounded patch in a given particle increases the probability of observing another unbounded patch in the same particle, a kind of correlation which is not contained in an independent bonding process. Since the potential energy cost for breaking a bond is, to a great extent, independent on the bonding environment, this suggests a significant role of the entropic component in the bonding free energy. Thus, for instance, if the weakly bounded particles localize themselves, the entropic gain when breaking a bond of one of these particles would be greater than the entropic gain when breaking a bond of a fully bounded particle. This entropic contribution enhances the population of the weakly bounded particles (in particular of the monomers) with respect to that expected for an independent bond-breaking process. In simple macroscopic terms, since both the binomial distribution of bonds given by Equation (5) and that obtained from the simulation have the same potential energy, \textit{i.e.}, they come from the same $p_b$ (or ${\cal N}_b$) value, and since the system in equilibrium at any $T$, the entropic gain must be the only reason for the deviation. Thus, as compared with an independent bonding process, the system minimizes its free energy by allowing more particles to be unbounded. However, and since both the binomial distribution and that obtained from the simulation have the same $p_b$ (or ${\cal N}_b$) value, the deviation in the population of monomers (and, in general, in the weakly bounded particles) must be balanced by the deviations in the populations of the almost fully bounded particles which (according to our previous microscopic interpretation) would be also more localized than what would be expected for an independent distribution of bonds. This microscopic picture, \textit{i.e.}, the localization of the particles according to their number of bonds (a kind of spatial bonding heterogeneity), will play a central role in our picture of the microscopic scenario describing the structure and dynamics of the system.\\
  
\subsection{Dynamics}

The dynamic behavior of this model has been partially investigated in Ref.~\cite{rovigatti_molphys}. Here we explore the individual and collective dynamics of the system in real and Fourier space. In particular we thoroughly investigate the connection between dynamics and system connectivity. Special attention is paid to the emergence of non-Gaussian distributions of displacements at low temperature. Finally, we concentrate on the self and collective mechanisms that govern the relaxation dynamics of the system within a wide range of spatial scales to finally propose a coherent microscopic picture for the low $T$ liquid. \\

\subsubsection{Mean Square Displacement}

\begin{figure}[tb]
\center
\includegraphics[width=0.76\linewidth]{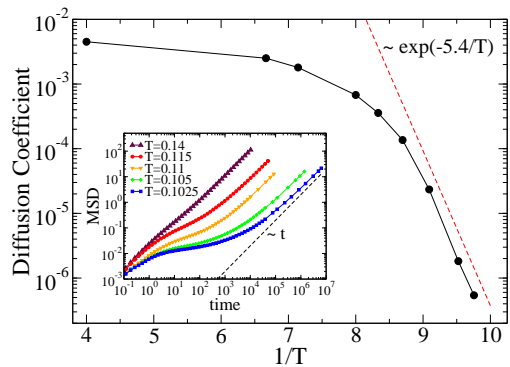}
\caption{Diffusion Coefficient as a function of $1/T$ (black line with solid circles). The figure also shows an Arrhenius law $e^{-5.4/T} \sim (1-p_b)^4 \sim (e^{-1.35/T})^4$ (dashed line). Inset: Mean square displacement as a function of time at different temperatures.}
\center
\label{Fig_Diffusion_MSD}
\end{figure}

Fig.~\ref{Fig_Diffusion_MSD} shows the $T$-dependence of the diffusion coefficient, $D$, estimated via the long time limit of the mean square displacement (MSD) (inset in Fig.~\ref{Fig_Diffusion_MSD}). Interestingly, $D$ becomes Arrhenius at low $T$ with an activation energy of about $5.4$. We should note that despite diffusivity was already studied for this model in Ref.~\cite{rovigatti_molphys}, the value reported here for the activation energy is larger than the one obtained previously by almost $20\%$ ~\cite{rovigatti_molphys}. Such difference is due to the much more lengthy simulations performed here, which led to sensibly improved statistics for all investigated quantities. This activation energy coincides with the activation energy of $(1-p_b)^4$ (see Fig.~\ref{Fig_Connectivity}b)) as previously reported in other models of tetrahedrally bounded particles~\cite{fully_bonded_tetra}. This result was previously interpreted as evidence that diffusion was completely dominated by the monomers. However, in this model the $T$-dependence of $(1-p_b)^4$ does not correspond to the $T$-dependence of the population of monomers, ${\cal N}_0$ (see Fig.~\ref{Fig_Connectivity}b)). As a consequence, the understanding of the relation $D(T) \sim (1-p_b(T))^4$, which links the long time dynamics of the system to a purely static observable, is still elusive. Nevertheless, as we next show, the different $T$-dependence of  $(1-p_b)^4$ and ${\cal N}_0$ does not exclude the significance on the total dynamics due to the population of monomers (and in general that of the weakly bounded particles).\\ 

\begin{figure}
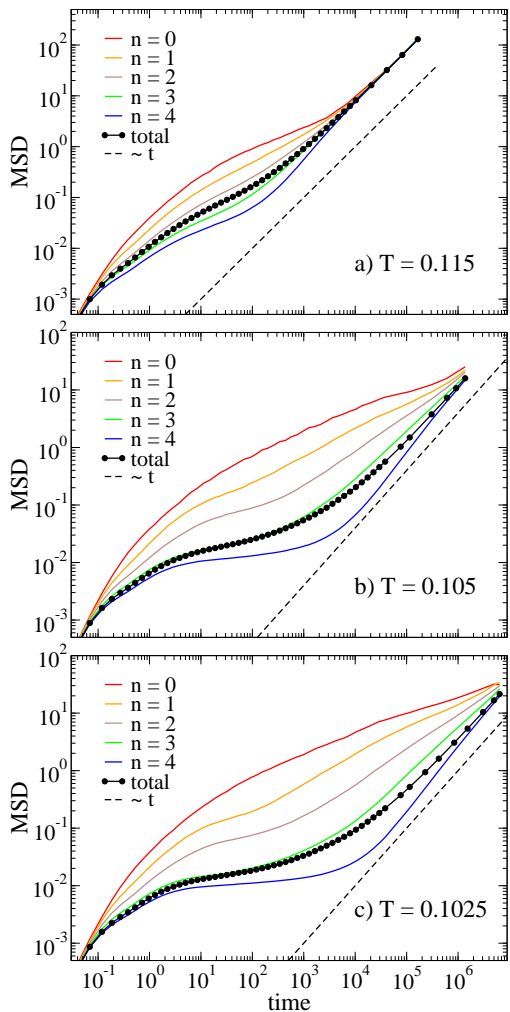

\includegraphics[clip,width=0.37\textwidth]{MSD_T0.115.eps}
\includegraphics[clip,width=0.37\textwidth]{MSD_T0.105.eps}
\includegraphics[clip,width=0.37\textwidth]{MSD_T0.1025.eps} \\
\caption{MSD as a function of time at different temperatures and for the different populations of particles discriminated by their number of bonds, $n$, at $t=0$. Each figure also incorporates the corresponding total MSD (black lines with black circles) as in the inset of Fig.~\ref{Fig_Diffusion_MSD}. Dashed lines represent the asymptotic diffusive behavior ($<(\Delta \vec{r}(t))^2> \sim t$).}
\label{Fig_MSD_Bond}
\end{figure}

To deepen in the discussion advanced in the previous paragraph we separate the total MSD into different populations. We have already seen how for this system we can discriminate between different particle populations, where each population includes all the particles with a common number of bonds. This is the case of Fig.~\ref{Fig_MSD_Bond}, where we present the MSD for different populations at different $T$'s, where each population has a common $n$ ($\in{\lbrace0,1,2,3,4\rbrace}$) at $t=0$. In this respect, and according to Equation (4), a given particle is considered to have $n$ bonds when its potential energy $E \in (-n,-n+1)$ (formally, particles with $n = 0$ have $E \in (0,\infty)$ since the repulsive part of the potential energy, Equation (2), is not a pure hard-sphere interaction). We note that, due to their evolving bonding state, all the particles with a given (initial) bonding state will change their number of bonds with time, eventually covering all the possible bonding states ($n$ $\in{\lbrace0,1,2,3,4\rbrace}$) to ensure ergodicity. Thus, by this discrimination we follow the system from its short time dynamics (where the initial bonding state is still present) to the long time dynamics (where the initial bonding state is not present anymore). Fig.~\ref{Fig_MSD_Bond} shows how at short times, and for any $T$, the MSD increases more rapidly upon decreasing the initial number of bonds, showing a short time window where ballistic effects are relevant for the weakly bounded particles. At intermediate times the separation (the spread of the curves) of the MSD for the different populations increases upon cooling the system as a clear sign of a dynamical heterogeneity. Thus, the weakly bounded particles move in general larger distances than those strongly bounded before reaching the intermediate plateau. This plateau is typically reached at those times at which the weakly bounded particles start losing the memory of their initial bonding state (see Ref.~\cite{rovigatti_molphys} for details on the bond lifetime scales of the system). In particular, at low $T$ (Fig.~\ref{Fig_MSD_Bond}c)) and for intermediate times the difference in the height of the plateau is significant when we go from $n = 2$ to $n = 3$. However from $n = 3$ to $n = 4$ the difference is relatively small pointing out that 3 bonds are probably sufficient to already arrest a particle. For the sake of clarity, we should notice that here we are discussing the dynamical heterogeneity associated to the populations of particles with a different number of bonds and not a real spatial dynamical heterogeneity. Indeed, a spatial dynamical heterogeneity, which could also be present in the system, would not be detected by the MSD since this observable does not include information on local spatial correlations.\\

At long times, the different MSD's converge to a common curve: the total MSD (black lines with black circles in Fig.~\ref{Fig_MSD_Bond}).  This convergence typically happens at the time at which all the populations have almost completely lost the memory of their initial number of bonds, \textit{i.e.}, at those times of the order of the average bond lifetime (see again Ref.~\cite{rovigatti_molphys} for details on the bond persistence). In this respect, it is interesting to notice how at long times, but before reaching the convergence, those particles that were strongly bounded at $t=0$ show an apparent super-diffusive motion which is analogous to that shown by the weakly bounded particles at short times (see for example blue and green lines in Fig.~\ref{Fig_MSD_Bond}b) and c)). This regime would correspond to those times at which a significant fraction of the initially fully bounded particles have already passed through a weakly bounded state and perform a large displacement. Reciprocally,  for this time scale, the initially weakly bounded particles show a sub-diffusive regime that would correspond to their pass through a tightly bounded state (see for example red and orange lines in Fig.~\ref{Fig_MSD_Bond}b) and c)).\\

Concerning the total MSD, we also see two effects. First, upon cooling the system the total MSD curve moves towards the partial MSD corresponding to the particles with four bonds. This is obviously due to the fact that ${\cal N}_4 \rightarrow 1$ upon cooling the system, leading the system to a complete arrest where the only remaining population to construct the total MSD is that corresponding to the 4-bounded particles. In addition, we anticipate a subsequent discussion by pointing out that despite at low $T$, and for the longest time investigated, the partial MSD's have not yet completely converged to the total MSD (Fig.~\ref{Fig_MSD_Bond}c)), the total MSD seems to reach the diffusive regime from which we extracted the diffusive coefficient (Fig.~\ref{Fig_Diffusion_MSD}).\\

\subsubsection{van Hove Function}

Since the MSD does not contain the total dynamic information of the system (it merely represents the time evolution of the variance of the total distribution of displacements), in systems with non-trivial dynamics it is worthwhile to explore the total distribution of the individual particle displacements through the Self van Hove function, $G_s(\vec{r},t)$~\cite{hansen}: 

\begin{equation}
G_s(\vec{r},t) = \frac{1}{N} \left\langle \sum_{i=1}^{N} \delta [\vec{r} - \Delta\vec{r}_i(t)] \right\rangle 
\end{equation}
\\

\noindent  
Where $G_s(\vec{r},t)$ represents the fraction of particles which have performed a given displacement $\Delta\vec{r}_i(t) = \vec{r}_i(t) - \vec{r}_i(0) = \vec{r}$ in a time $t$. For a pure Gaussian distribution of displacements $G_s(r,t)/r^2 \sim e^{-r^2/6Dt}$ at any time (note that here we directly consider a radial displacement $r$ and therefore we normalize by $r^2$ since we have an isotropic three-dimensional system).\\

\begin{figure}[tb]
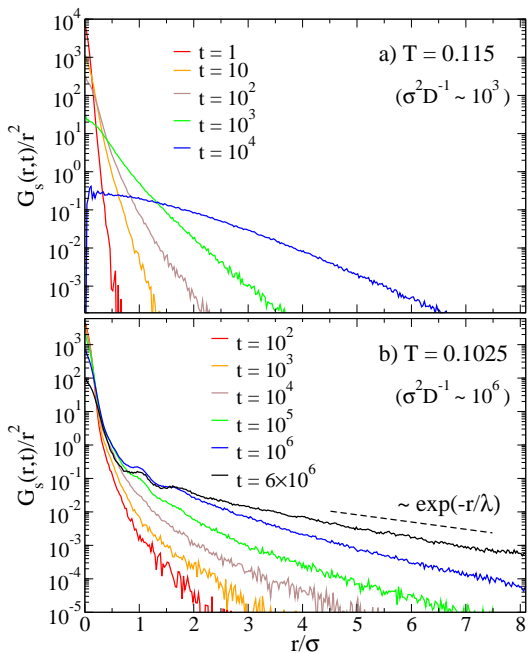

\includegraphics[clip,width=0.8\linewidth]{self_van_Hove_T0.115.eps}
\includegraphics[clip,width=0.8\linewidth]{self_van_Hove_T0.1025.eps} \\
\caption{Normalized Self van Hove function $G_s(r,t)/r^2$ for different times at a) T=0.115 and b) T=0.1025. The figure also shows for each $T$ a characteristic time $\sigma^2D^{-1}$ as an indication of the emergence of the diffusive regime. In b) we also show an exponential decay with $\lambda / \sigma = 2$ (dashed line).}
\label{Fig_self_van_Hove}
\end{figure}

Departures from the Gaussian behavior of the Self van Hove function have already been reported in the literature for different systems~\cite{pinaki_non_gaussian,hs_non_gaussian} where a clear indication of the non-Gaussian dynamics is manifested by the emergence of a broad exponential tail in $G_s(r,t)/r^2 \sim e^{-r/ \lambda (t)}$, where $\lambda (t)$ is a characteristic length that increases with time. In Fig.~\ref{Fig_self_van_Hove} we show $G_s(r,t)/r^2$ for two different temperatures which correspond to the beginning of the Arrhenius behavior ($T=0.115$, see Fig.~\ref{Fig_Diffusion_MSD}) and to the lowest temperature investigated ($T=0.1025$). In particular, Fig.~\ref{Fig_self_van_Hove}a) shows three different time regimes. At very short times ($t \simeq 1$), \textit{i.e.}, times which are significantly smaller than the cage residence time, $G_s(r,t)/r^2$ is peaked around the origin, describing the vibrational disorder with a behavior which is not far from a Gaussian distribution. At intermediate times  ($t \simeq 10^1-10^3$), rare events appear in the form of large individual displacements and $G_s(r,t)/r^2$ shows a clear exponential tail with a characteristic length which increases with time. In this regime only a small fraction of particles has performed jumps while the majority of the particles are still vibrating in their local environment. Only when all the particles have moved significantly, \textit{i.e.}, for times typically longer than the residence time ($t \gtrsim 10^4$), Gaussian behavior is recovered. This time almost coincides with the time at which motion is already diffusive (Fig.~\ref{Fig_MSD_Bond}a)).\\

Fig.~\ref{Fig_self_van_Hove}b) shows $G_s(r,t)/r^2$ for the lowest $T$ investigated. In this case, the final Gaussian regime is still not observed within the time of our simulation and only the first two regimes can be detected. Indeed, at long times the correlation length of the exponential decay of $G_s(r,t)/r^2$ is of the order of few particle diameters, $\lambda (t_{long}) / \sigma \cong 2$. Apart from the long range exponential behavior, at long times  we can also appreciate some undulations in $G_s(r,t)/r^2$ at distances around $r / \sigma \cong 1$ and $r / \sigma \cong 1.7$ which correspond to the first and second peaks of the structure factor, $S(q)$, where $q_{peak} \cong 2 \pi / r$ (see $T = 0.1025$ in the Appendix, Fig.~\ref{Fig_Sq_total}). These undulations reflect how the tetrahedral structure of the system at low $T$ favors those displacements corresponding to the nearest and tetrahedral neighbor distances.\\

It is interesting to notice that at low $T$ and for the longest time investigated ($t \cong 10^7$) motion seems to be already diffusive (see the total MSD in Fig.~\ref{Fig_MSD_Bond}c) and the inset in Fig.~\ref{Fig_Diffusion_MSD}). However, as mentioned in the previous paragraph, the distribution of displacements is non-Gaussian. Recently this diffusive but still non-Gaussian behavior has aroused interest, it is the so-called "\textit{anomalous yet Brownian
diffusion}"~\cite{pnas_non_gaussian_granick,nat_mat_non_gaussian_granick,hs_non_gaussian}. There have been some attempts to rationalize this anomalous diffusion but all of them rely on a description of the heterogeneous dynamics of the system with no explicit consideration of real \textit{spatial} heterogeneities~\cite{pinaki_non_gaussian,diffusing_diffusivity}. In particular, the problem of whether or not this anomalous diffusion can be understood in terms of an underlying heterogeneous structure remains unclear. Nevertheless we should take into account that the observation of the non-Gaussian behavior will, in principle, depend on the time observational window. In other words, the Gaussian behavior will be recovered as soon as the hypotheses of the Central Limit Theorem hold~\cite{gardiner}. In this respect, the distribution of displacements will be Gaussian when it results from the average of independent and equally distributed particle displacements. However, at low $T$ and for the longest time investigated we have already pointed out that the system still presents differences in the dynamics between the populations of particles according to their initial number of bonds. For instance, although at long times the total MSD seems to be linear in time, the partial MSD's presented in Fig.~\ref{Fig_MSD_Bond}c) have not yet converged, pointing out that the distribution of displacements of the different populations are not equally distributed over our time window. \\

\begin{figure}[tb]
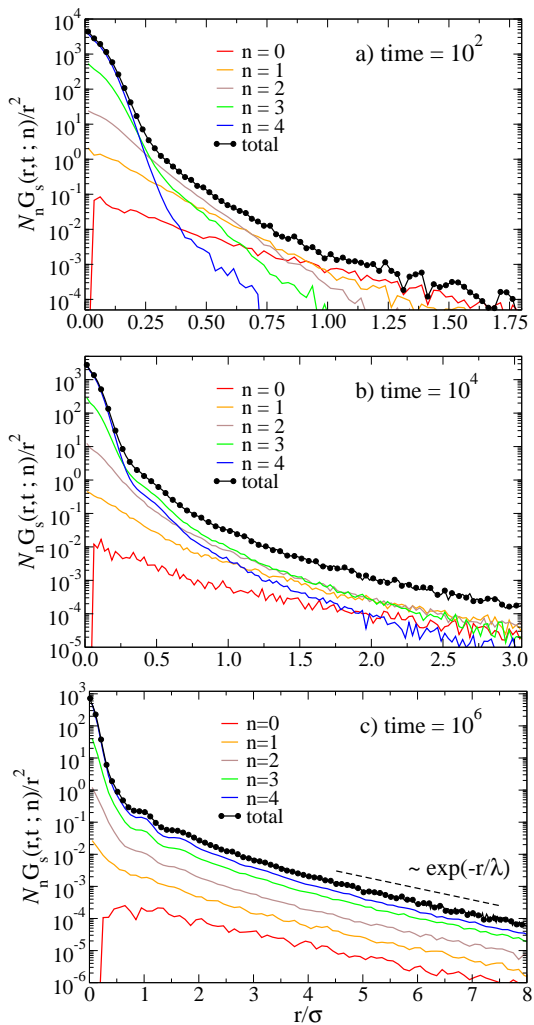

\includegraphics[clip,width=0.8\linewidth]{self_van_Hove_T0.1025_t1e2_bond.eps} \vspace{0.3cm}

\includegraphics[clip,width=0.8\linewidth]{self_van_Hove_T0.1025_t1e4_bond.eps} \vspace{0.15cm}

\includegraphics[clip,width=0.8\linewidth]{self_van_Hove_T0.1025_t1e6_bond.eps}
\caption{Partial Self van Hove functions ${\cal N}_nG_s(r,t ;n)/r^2$ at $T=0.1025$ for the different populations of particles according to their initial number of bonds, $n$ ($\in{\lbrace0,1,2,3,4\rbrace}$). a) $t=10^2$, b) $t=10^4$, and c) $t=10^6$. The figure also shows the corresponding $G_s(r,t)/r^2$ for the total number of particles (black lines with solid circles). In c) we also show an exponential decay with $\lambda / \sigma = 1.2$ (dashed line) which is slightly different to that present in Fig.~\ref{Fig_self_van_Hove}b) since there is a difference between the longest time in this figure and that corresponding to Fig.~\ref{Fig_self_van_Hove}b) ($t = 10^6$ and $t = 6 \times 10^6$ respectively).}
\label{Fig_self_van_Hove_bond}
\end{figure}

As done for the MSD, we have also separated the Self van Hove function into different populations. Thus Fig.~\ref{Fig_self_van_Hove_bond} shows, for different times and for the lowest $T$ investigated, the contributions to the total Self van Hove function due to the different populations of particles according to their number of bonds at $t = 0$. We should note that in order to properly compute the contribution of each population to the total Self van Hove function we should re-scale the Self van Hove function of each population by its corresponding fraction of particles, ${\cal N}_n$ ($n \in{\lbrace0,1,2,3,4\rbrace}$). Thus we represent ${\cal N}_nG_s(r,t ;n)/r^2$, where the total Self van Hove function results $G_s(r,t)/r^2= \sum_{n=0}^4 {\cal N}_nG_s(r,t ;n)/r^2$. Again, at any time and for short displacements the main contribution to $G_s(r,t)/r^2$ is due to ${\cal N}_4G_s(r,t ;n=4)/r^2$ (see blue lines in Fig.~\ref{Fig_self_van_Hove_bond}) since for short displacements we are just computing vibrational motion and therefore the main contribution should arrive from the biggest population which at low $T$ is that corresponding to the 4-bounded particles. At short times and for long displacements (Fig.~\ref{Fig_self_van_Hove_bond}a)) we see how the main contribution to $G_s(r,t)/r^2$ is mainly due to the monomers (${\cal N}_0G_s(r,t ;n=0)/r^2$) or, in general, to the weakly bounded particles (red line in Fig.~\ref{Fig_self_van_Hove_bond}a)). This result was already anticipated in the discussion concerning the MSD for the different populations where we already noticed the short time super-diffusive regime of the weakly bounded particles (Fig.~\ref{Fig_MSD_Bond}c)).  At intermediate times (Fig.~\ref{Fig_self_van_Hove_bond}b)), particles start to lose their initial bonding state and the long displacement contribution to the total Self van Hove function is mainly due to those particles which had an intermediate numbers of bonds at $t = 0$ and which now have presumably passed through a weakly bonding state (green and orange lines in Fig.~\ref{Fig_self_van_Hove_bond}b)). Finally, at long times (Fig.~\ref{Fig_self_van_Hove_bond}c)), and when most of the particles have lost the memory of their initial number of bonds, we see how the main contribution (for short and long displacements) is mostly due to the biggest population, $n=4$ (blue line in Fig.~\ref{Fig_self_van_Hove_bond}c)). Contrary to short and intermediate times, where each population shows its own slope for the exponential tail at long distances, at long times (and once most of the particles have lost the memory of their initial number of bonds) all the populations present a clear exponential tail with an almost common slope ($\lambda (t_{long}) / \sigma \cong 1.2$, Fig.~\ref{Fig_self_van_Hove_bond}c)) .\\

 Once we have studied the distributions of the individual particle displacements through the Self van Hove function, we now consider the collective dynamics of the system in real space through the Distinct van Hove function, $G_d(\vec{r},t)$~\cite{hansen}:

\begin{equation}
G_d(\vec{r},t) = \frac{1}{N} \left\langle \sum_{i = 1}^{N} \sum_{j \neq i}^{N} \delta [\vec{r} - \vec{r}_j(t) + \vec{r}_i(0)] \right\rangle 
\end{equation}
\\

\begin{figure}
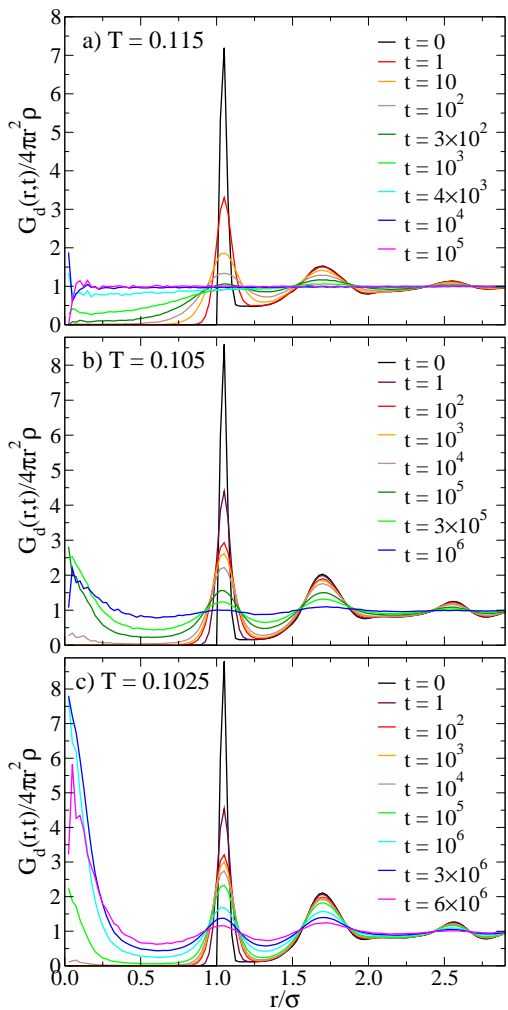

\includegraphics[clip,width=0.37\textwidth]{Dist_van_Hove_T0.115.eps}
\includegraphics[clip,width=0.37\textwidth]{Dist_van_Hove_T0.105.eps}
\includegraphics[clip,width=0.37\textwidth]{Dist_van_Hove_T0.1025.eps} \\
\caption{Normalized Distinct van Hove function $G_d(r,t)/4\pi r^2\rho$ at different temperatures and for several times.}
\label{Fig_Distinct_van_Hove}
\end{figure}

\noindent  
Where $G_d(\vec{r},t)$ counts all those correlations between particle $i$ at $t = 0$ and particle $j$ at $t \geq 0$ ($\forall i \neq j $) which are compatible with a given $\vec{r}$. Again, as previously done for the Self van Hove function, we present in  Fig.~\ref{Fig_Distinct_van_Hove} $G_d(r,t)/4\pi r^2\rho$ at different temperatures, where we normalize by $4\pi r^2\rho$ to compute the correlations according to the radial coordinate $r$ by also adding the number density to have $G_d(r,t=0)/4\pi r^2\rho \equiv g(r)$, being $g(r)$ the radial distribution function~\cite{hansen}. Thus at any $T$ and for $t=0$ we just have the corresponding radial distribution function which shows the tetrahedral structure previously discussed, where we see two main peaks at $r / \sigma \cong 1$ and $r / \sigma \cong 1.7$ which are related (according to a simple Bragg's interpretation) to the two main peaks in $S(q)$ (Fig.~\ref{Fig_Sq_total}) via $q_{peak} \cong 2 \pi / r_{peak}$.\\

The most interesting effect arrives upon cooling the system. Whereas at the beginning of the Arrhenius regime (Fig.~\ref{Fig_Distinct_van_Hove}a)) the system completely loses the memory of its dynamic correlations, \textit{i.e.}, $G_d(r,t_{long})/4\pi r^2\rho \cong 1$ $\forall r$, at low $T$ (Fig.~\ref{Fig_Distinct_van_Hove}b) and c)) we see how the system still presents a remaining correlation even for the longest simulation time. Thus  at long times and low $T$, and despite the double peak structure has almost disappeared, a significant fraction of particles still occupies the position that was occupied by other particles at $t=0$ (a typical feature of systems where hopping motion is present), \textit{i.e.}, $G_d(r,t)/4\pi r^2\rho$ increases as $r \rightarrow 0$. In particular at long times and for the lowest $T$ (Fig.~\ref{Fig_Distinct_van_Hove}c)), $G_d(r \rightarrow 0,t)/4\pi r^2\rho > 1$, that is, part of the dynamic correlations still survive despite most of the particles have individually moved a significant distance (see Fig.~\ref{Fig_MSD_Bond}c)). Certainly the system starts to lose the remaining memory for the longest simulation time (pink curve in Fig.~\ref{Fig_Distinct_van_Hove}c)). However, despite $G_d(r \rightarrow 0,t)/4\pi r^2\rho$ starts to decrease it is still far from the complete ergodic behavior $G_d(r,t_{long})/4\pi r^2\rho \cong 1$ ($\forall r$) which is indeed reached at intermediate temperatures (Fig.~\ref{Fig_Distinct_van_Hove}a)). Indeed, the system shows a significant rigidity at low $T$ (accounted for by the height of the maximum of $G_d(r \rightarrow 0,t)/4\pi r^2\rho$) compared with other strong glass forming-liquids~\cite{horbach_walter_silica} which is presumably due to the highly directional interaction (\textit{i.e}, to the small angular bond opening).\\ 

Before concluding this section, we should note that the separation of  the Distinct van Hove function into particle populations, as previously presented for the MSD and the Self van Hove function, does not have a straightforward interpretation. The Distinct van Hove function connects different particles at different times, therefore, missing the information of the number of bonds of the two correlated particles at a common time due to the evolving bonding state of the particles. Thus, a systematic study of this function would consider all the possible combinations between populations of particles with $n$ bonds at $t=0$ and $m$ bonds at $t \geq 0$ ($n,m \in{\lbrace0,1,2,3,4\rbrace}$).\\
 
\subsubsection{Non-Gaussian Parameter}
  
\begin{figure}[tb]
\includegraphics[width=0.9\linewidth]{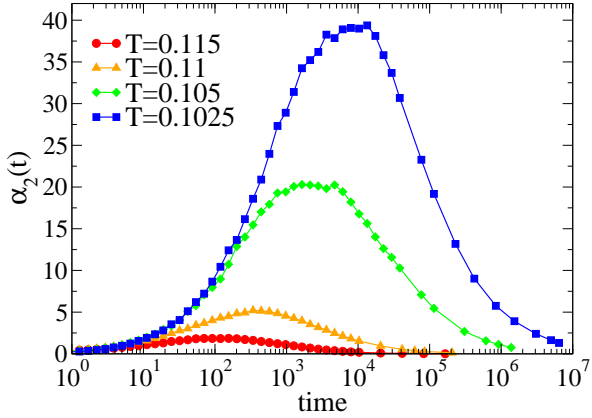}
\caption{Non-Gaussian parameter $\alpha_2(t)$ as a function of time at different temperatures.}
\label{Fig_alpha_2}
\end{figure}

An alternative procedure to determine and quantify the non-Gaussian statistics associated to the dynamics of a system at a given time relies on the estimation of the so-called non-Gaussian parameter, $\alpha_2(t)$~\cite{rahman_liquid_argon}:

\begin{eqnarray} 
\alpha_2(t)=\frac{3}{5}\frac{\langle (\Delta\vec {r}(t))^4\rangle}{\langle (\Delta \vec {r}(t))^2\rangle ^2}-1 
\end{eqnarray}
\\
\noindent
Where $\langle (\Delta\vec {r}(t))^4\rangle$ and $\langle (\Delta\vec {r}(t))^2\rangle$ (\textit{i.e.}, the MSD) are respectively the fourth and second moments of the distribution of displacements at time $t$ in three-dimensional space. For a Gaussian distribution $\alpha_2(t)=0$ whereas non-Gaussian distributions are manifested through a positive value of $\alpha_2(t)$. Fig.~\ref{Fig_alpha_2} shows $\alpha_2(t)$ at different temperatures. The three different regimes documented by the Self  van Hove function are also manifested in $\alpha_2(t)$. First, at short times, $\alpha_2(t) \cong 0 $ indicating that the distribution of the displacement is almost Gaussian. At intermediate times, we reach a maximum that indicates the time at which the distribution of displacements is farthest from the Gaussian behavior. The position and height of this maximum increases upon cooling the system, reaching values at low $T$ which are significantly higher than those reported for both fragile and  strong glass forming-liquids such as binary mixtures of Lennard-Jones particles~\cite{walter_andersen}, models of supercooled water~\cite{sciortino_sup_water}, or viscous silica~\cite{horbach_walter_silica_2} but comparable to other recently reported in network liquids with competing gel-glass phases~\cite{pinaki_walter_pablo_competing_gel}. Finally, at long times, $\alpha_2(t)$ decreases, and the distribution seems to recover the Gaussian statistics, $\alpha_2(t)=0$. However, we should mention that this contrasts with our previous discussion concerning the Self van Hove function for which the system at low $T$ and long times  still presented a clear exponential tail, therefore, not having recovered completely the Gaussian statistics. In this respect, we should notice that the Self van Hove function captures the whole distribution of displacements being a better estimator of the non-Gaussian statistics than $\alpha_2(t)$, which merely checks the relation between the second and fourth moments that would be expected for a Gaussian distribution.\\

\begin{figure}[tb]
\includegraphics[clip,width=0.9\linewidth]{alpha_2_T0.115_n.eps}\vspace{0.3cm}
\includegraphics[clip,width=0.89\linewidth]{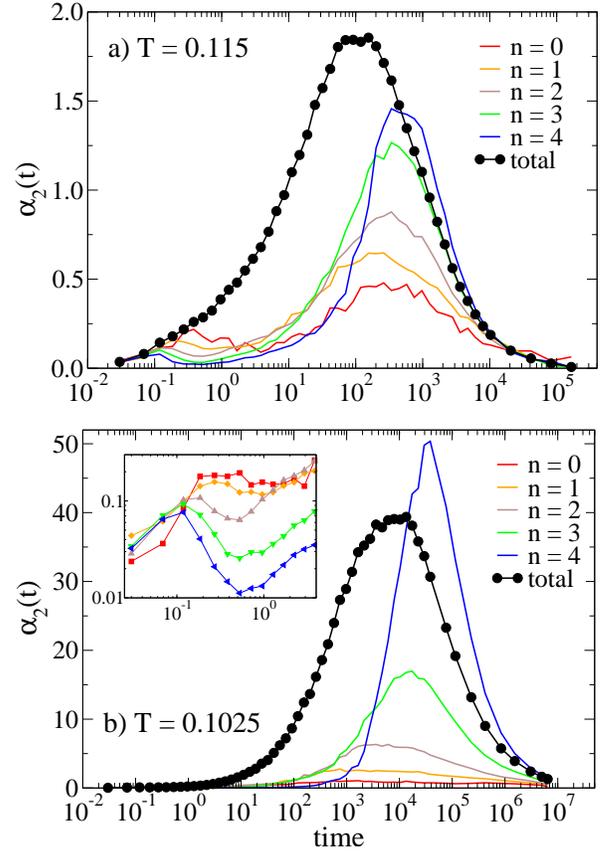}\\
\caption{Non-Gaussian parameter $\alpha_2(t)$ at different temperatures, a) $T = 0.115$ and b) $T = 0.1025$. Here $\alpha_2(t)$ is separated into different populations of particles according to their initial number of bonds, $n$ ($\in{\lbrace0,1,2,3,4\rbrace}$). The figure also shows the corresponding $\alpha_2(t)$ for the total number of particles (black lines with solid circles). Inset in b) shows in a double log plot a detail of the maxima corresponding to the local short time behavior of the particles separated by their initial number of bonds (colors in the inset are as in the main figure whereas symbols have been introduced as a guide for the eyes).}
\label{Fig_alpha_2_n}
\end{figure}

To gain intuition on the microscopic mechanism, we present in Fig.~\ref{Fig_alpha_2_n} $\alpha_2(t)$ separated into different populations of particles according to their number of bonds at $t = 0$, as previously done for the MSD and the Self van Hove function. Thus Fig.~\ref{Fig_alpha_2_n} shows $\alpha_2(t)$ for the different populations at the beginning of the Arrhenius regime ($T=0.115$, Fig.~\ref{Fig_alpha_2_n}a)) and for the lowest $T$ investigated ($T=0.1025$, Fig.~\ref{Fig_alpha_2_n}b)). At intermediate $T$ (Fig.~\ref{Fig_alpha_2_n}a)) we see how the main maxima corresponding to the different populations almost appear at a common time ($t \approx 10^3$), suggesting that the time needed for all populations to disseminate the memory of their initial bonding state (and therefore their initial degree of mobility) is almost similar. This time is also similar to the bond lifetime reported in Ref.~\cite{rovigatti_molphys} for the same temperature. However, at low $T$ (Fig.~\ref{Fig_alpha_2_n}b)) the maxima corresponding to each population appear later for those particles that are more tightly bounded at $t = 0$, despite the order of the time associated to the total $\alpha_2(t)$ is still similar to the bond lifetime for this temperature~\cite{rovigatti_molphys}. In this respect, we can speculate with the idea that at low $T$ there is difference between the bond lifetime of the different populations. We also point out that the total $\alpha_2(t)$ (Fig.~\ref{Fig_alpha_2_n}a) and b)) has a maximum that occurs earlier than the partial $\alpha_2(t)$ associated to the different populations, an effect which is related to the way all the populations are mixed when performing the average to obtain the total $\alpha_2(t)$. Also interesting is the effect which is present at short times in the form of a less obvious maximum. These short time maxima, which are fairly insensitive to temperature, have been enhanced in the inset of Fig.~\ref{Fig_alpha_2_n}b) since their relative height with respect to the main maxima is too small (short time maxima at $T=0.115$, Fig.~\ref{Fig_alpha_2_n}a), are directly observable). These short time maxima increase their height upon decreasing $n$, reflecting that the intra-cage motion of the particles is ``less Gaussian'' for the free particles than for the tightly bounded particles. This suggests that the free particle local environment is more heterogeneous and, therefore, the short time displacements of the weakly bounded particles are "less equally distributed".\\

\begin{figure}[tb]
\includegraphics[width=0.85\linewidth]{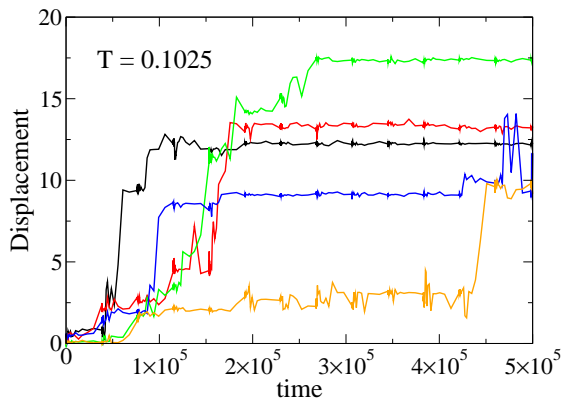}
\caption{Displacement of some selected jumping particles as a function of time at $T=0.1025$. The particles perform sporadic large displacements over short time intervals while most of the time remain vibrating. Each color represents the displacement of one selected particle. We highlight that, contrary to the previous figures, here the time axes is linear in order to clearly illustrate the real interval duration.}
\label{Fig_Disp_jumpers}
\end{figure}

We finally illustrate the rare event dynamics present in the system at low $T$ by showing in Fig.~\ref{Fig_Disp_jumpers} the displacement of some selected particles (\textit{jumpers}) which clearly show their intermittent dynamics manifested through sporadic large displacements (jumps) at the single particle level. Although not statistically relevant, the figure indeed suggests that the single particle motion can be described in terms of a vibrational (stationary) dynamics (where the particle is vibrating in a confining cage created by their neighbors) which is interrupted by large jumps whose time duration is significantly smaller than the time spent during the local vibrations. Fig.~\ref{Fig_Disp_jumpers} also shows that during the jumps, the selected particles move over distances of the order of few particle diameters.

\subsubsection{Scattering Functions: Cooperative Domains}
           
Next we present the study of the relaxation time at different spatial scales for both Self and Collective scattering functions~\cite{hansen} (some selected low-$T$ correlation functions are presented in the Appendix, Fig.~\ref{Fig_f_self_coll_T0.1025}). The large box size of our simulation (see Section II) allows us to reach wavevectors of $q\sigma= 2\pi/L \cong 0.25$, that is, more than one order of magnitude smaller than the inverse of the nearest neighbor distance. We measure the corresponding relaxation times associated to the Self and Collective scattering functions, $\tau_{self}(q)$ and $\tau_{coll}(q)$, by fitting the final $\alpha$-decay of the corresponding correlation function by means of a stretched exponential function~\cite{hansen} (see Appendix). By analogy with the Distinct van Hove function, here we study those correlations included in the Collective scattering function for the total set of particles and, therefore, we do not discriminate into particle populations. We also note that similar relaxation times as those presented here were reported in Ref.~\cite{rovigatti_molphys}. However, we are now able to better understand the underlying physics behind the current results since the data reported herein have been obtained through much lengthier simulations, providing a more solid ground for the interpretation that follows.\\

\begin{figure}[tb]
\center
\includegraphics[width=0.9\linewidth]{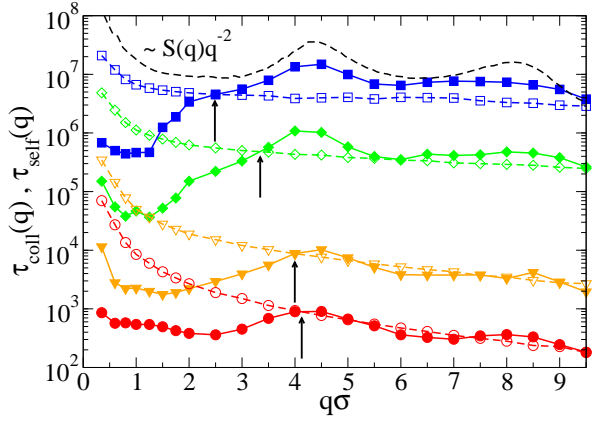}
\caption{Relaxation times corresponding to the Collective (full symbols), $\tau_{coll}(q)$, and the Self (empty symbols), $\tau_{self}(q)$, scattering functions as a function of $q$ at different temperatures ($T=0.1025$ - blue squares, $T=0.105$ - green diamonds, $T=0.11$ - orange triangles, and $T=0.115$ - red circles). Vertical arrows mark the crossing between $\tau_{self}(q)$ and $\tau_{coll}(q)$ at $q^{*}(T)$ ($\sim \xi(T)^{-1}$). Also included in the figure is a $\gamma q^{-2}S(q)$ trend (black dashed line), where we took the $S(q)$ corresponding to $T = 0.1025$ and fixed the prefactor $\gamma$ to have $\gamma q^{-2}S(q) = \tau_{coll}(q)$ at large $q$.}
\center
\label{Fig_tau_self_coll}
\end{figure}

We first present in Fig.~\ref{Fig_tau_self_coll} the comparison between $\tau_{self}(q)$ and $\tau_{coll}(q)$ as a function of $q$.  At large $q$ (wavevectors typically greater than the main peak of the structure factor) both $\tau_{self}(q)$ and $\tau_{coll}(q)$ follow the expected common trend (for any $T$) by converging to a common curve (this trend has been checked in this system for  values as large as $q\sigma = 30$). Also at any $T$, $\tau_{self}(q)$  monotonically increases on decreasing $q$, showing that the relaxation time of the individual particle dynamics increases upon increasing the observational spatial scale. $\tau_{coll}(q)$ shows a non-monotonic behavior at intermediate $q$ whose oscillations are in phase with the structure factor (\textit{de Gennes narrowing}~\cite{pusey_1975,deGennes_1959}). However the complete behavior of $\tau_{coll}(q)$ within the explored $q$-range cannot be explained by a behavior in phase with the structure factor (see the departure in Fig.~\ref{Fig_tau_self_coll} between the $q^{-2}S(q)$ trend and $\tau_{coll}(q)$ for $T =0.1025$). At intermediate $q$ (values around the tetrahedral peak of the structure factor) $\tau_{self}(q)$ and $\tau_{coll}(q)$ start to decouple upon cooling the system. This effect can be followed by the increasing separation of $\tau_{self}(q)$ and $\tau_{coll}(q)$ which reaches near one order of magnitude at the position of the tetrahedral peak for the lowest $T$ investigated ($\tau_{coll}(q) \cong 10 \times \tau_{self}(q)$ at $T = 0.1025$ for $q\sigma = 4.5$). Since at low $T$ (and for intermediate wavevectors) $\tau_{coll}(q) > \tau_{self}(q)$, we can infer that the system maintains its dynamic collective correlations at distances even larger than that associated to the tetrahedral peak although the particles have individually moved even larger distances. Despite the system at low $T$ presents an amorphous tetrahedral structure, this phenomenology at intermediate distances reminds us the archetypal crystal behavior, where particles can diffuse despite the structure being permanent. As we next discuss, this \textit{long-lasting} correlation present in the system  will have associated a dynamic length scale of the order of the inverse of the $q$ value at which $\tau_{self}(q)$ and $\tau_{coll}(q)$ cross, that is, the spatial scale at which the relaxation time associated to the individual particle relaxation start to overpass that of the collective particle relaxation.\\

Interestingly, the crossing between $\tau_{self}(q)$ and $\tau_{coll}(q)$ in Fig.~\ref{Fig_tau_self_coll} appears at shorter $q$'s upon cooling the system, suggesting the emergence of a dynamic length scale associated to the collective dynamics which increases upon decreasing $T$ (vertical arrows in Fig.~\ref{Fig_tau_self_coll}). Thus, the crossing between $\tau_{self}(q)$ and $\tau_{coll}(q)$ at a given $q^{*}(T)$ allows us to define a dynamic length scale, $\xi(T)$, as $\xi(T) \sim q^{*}(T)^{-1}$. By a simple inspection of the left-shift of $q^{*}(T)$ upon decreasing $T$, we can conclude that $\xi(T)$ increases by almost a factor 2 within the explored $T$-range. However, more systematic work is needed to determine and rationalize a precise functional behavior for $\xi(T)$. We should note that similar low $T$-values (\textit{i.e.} values of the order of few particle diameters) have already been reported in the literature of glass-forming liquids for different dynamic length scales but they are typically based on different methodologies as that exposed here \citep{binder_walter,biroli_berthier}. For instance, multi-point dynamic susceptibilities have been used to estimate the size of dynamic cooperative regions \cite{berthier_science}, response of the dynamic structure factor to a external potential has been proposed to infer the existence of a diverging dynamic length scale \cite{biroli_MCT_length}, local pair correlators have been analyzed to prove the emergence of dynamic heterogeneities \cite{berthier_heterogeneities}, and point-to-set correlation functions have been used to detect long-range dynamic correlations \cite{wall}.\\

Appealingly, here the increasing dynamic length scale suggests a heuristic microscopic picture for the low $T$ regime where the system  would be conformed by cooperative domains of particles or, more precisely, by \textit{dynamically} correlated regions (CR). The $T$-dependent linear size of these CR's would be given by $\xi(T) \sim q^{*}(T)^{-1}$. Thus, upon cooling the system the CR's would become larger and the network liquid stiffer. The effect of the CR dynamics at low $T$ is that  $\tau_{coll}(q)$  remarkably drops by more than one order of magnitude for $q$ smaller than the CR length scale (typically $q\sigma \lesssim 2.5$ at $T = 0.1025$). Moreover, and in order to ensure the final ergodicity of the system, the set of particles defining the CR's will change with time (typically at a time of the order of the $\tau_{self}(q)$-$\tau_{coll}(q)$ crossing).\\
 
\begin{figure}[tb]
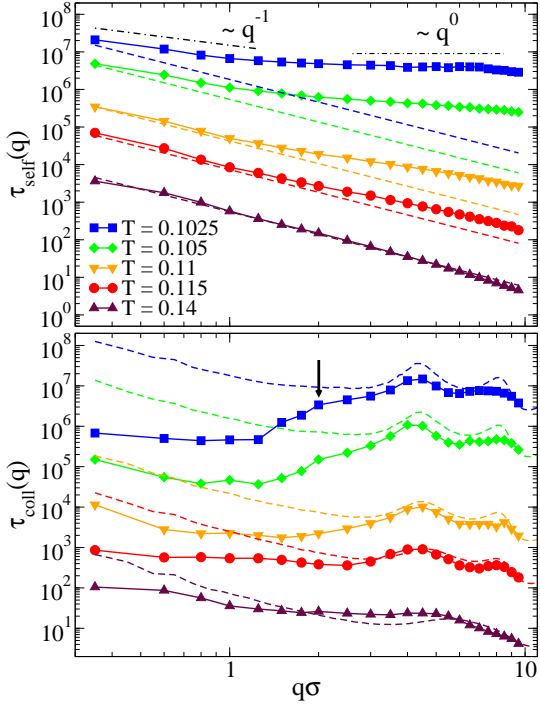

\center
\includegraphics[clip,width=0.82\linewidth]{tau_self.eps}
\includegraphics[clip,width=0.82\linewidth]{tau_coll.eps}
\caption{a) Log-log plot of the relaxation time corresponding to the Self scattering function, $\tau_{self}(q)$, as a function of $q$ at different temperatures. Dashed lines (with colors according to each $T$ ) represent the diffusive behavior $q^{-2}D^{-1}$ expected for a high $T$ liquid. At low $T$ we have also included a $q^{-1}$ trend at short $q$ and a constant $q^0$ trend at intermediate and large $q$ (black dotted-dashed lines), b) Log-log plot of the relaxation time corresponding to the Collective scattering function, $\tau_{coll}(q)$, as a function of $q$ at different temperatures (colors and symbols are as in a)). Also included in the figure is a $\gamma q^{-2}S(q)$ trend (dashed lines with colors according to each $T$), where we took the $S(q)$ corresponding to each $T$ and fixed the prefactor $\gamma$ to have $\gamma q^{-2}S(q) = \tau_{coll}(q)$ at large $q$. We highlight by a vertical arrow the emergence of a shoulder at $T = 0.1025$ whose presence does not appear in the Structure factor (see Fig.~\ref{Fig_Sq_total}).}
\center
\label{Fig_tau_pannel}
\end{figure}

Let us now discuss separately the behavior of $\tau_{self}(q)$ and $\tau_{coll}(q)$ to distinguish their different $q$-regimes at different $T$'s. Thus, Fig.~\ref{Fig_tau_pannel}a) shows $\tau_{self}(q)$ from the high $T$ regime ($T=0.14$) to the lowest $T$ investigated. At $T=0.14$ we have $\tau_{self}(q) \cong q^{-2}D^{-1}$ within the whole $q$-range, \textit{i.e.}, the behavior expected for a pure diffusive process. Upon cooling the system we see a clear departure of $\tau_{self}(q)$ from the diffusive behavior which is more pronounced at large $q$. Indeed, upon cooling the system  $\tau_{self}(q)$ tends to be $q$ independent at intermediate $q$ ($\sim q^0$) whereas for small wavevectors a $q^{-1}$-dependence appears (see  $q\sigma \lesssim 1.5$ at $T=0.1025$ in Fig.~\ref{Fig_tau_pannel}a)).\\

We first discuss the intermediate $q$-dependence of $\tau_{self}(q)$ at low $T$ (in particular at $T=0.1025$). At intermediate $q$, we see a clear departure of $\tau_{self}(q)$ from $q^{-2}D^{-1}$, where $\tau_{self}(q)q^2 D \sim 100$ at $q\sigma = 10$. Thus, at low $T$ and for intermediate spatial scales (typically $2 \lesssim q\sigma \lesssim 8$), the time needed to decorrelate the Self scattering function clearly exceeds the time $q^{-2}D^{-1}$, which is the time that all the particles would need to move a distance of the order of $2\pi/q$ if the process were diffusive with coefficient $D$. Since the system at low $T$ presents a hopping dynamics, decorrelation at intermediate $q$ requires that most of the particles perform at least one jump, thus leaving their local environment. Since these jumps are intermittent we should wait a common (constant) time for all intermediate $q$'s in order to perform the particles their jumps and escape from their local environment. In this respect, we should note that the value of $\tau_{self}(q)q^2 D \approx 100$ around the main peak of the structure factor provides us with an estimation of the so-called \textit{translational decoupling} ~\cite{pinaki_non_gaussian}, that is, the ratio between the time associated to the real intermittent dynamics present in the system at intermediate spatial scales (\textit{i.e.}, the mean cage residence time) and the time that would be expected from a \textit{continuous} diffusive dynamics at the same intermediate spatial scales. This translational decoupling appears to be significant in the present system at low $T$ compared with other values previously reported for fragile and strong glass forming-liquids~\cite{pinaki_non_gaussian}.\\

We now consider the interesting $q^{-1}$ behavior of $\tau_{self}(q)$ at low $T$ and short $q$. Before entering into the microscopic picture to interpret such as behavior we recover our previous discussion on the anomalous diffusion but restated in terms of the Self scattering function. In principle, for a Gaussian distribution of displacements the Self scattering function, $F_{self}(q,t)$, would read~\cite{davenport_root,paul_pusey}:

\begin{eqnarray} 
F_{self}(q,t)= \frac{1}{N} \left\langle \sum_{j=1}^{N} e^{\textit{i} \vec{q} \cdot \Delta \vec{r}_j(t)} \right\rangle = e^{-\frac{1}{6}q^2<(\Delta \vec{r}(t))^2>} \,\,\,\,\,\,\,\,\,\,\
\end{eqnarray}
\\
\noindent
If, in addition, we suppose that motion is diffusive ($<(\Delta \vec{r}(t))^2> = 6D t$) that would give us $\tau_{self}(q) \sim q^{-2}$ (which is indeed the result in Fig.~\ref{Fig_tau_pannel}a) at high $T$). As discussed in the MSD section, at low $T$ we reach the diffusive behavior at long times, however $\tau_{self}(q) \sim q^{-1}$ instead of $\tau_{self}(q) \sim q^{-2}$. Thus the only reason for not having the expected $\tau_{self}(q) \sim q^{-2}$ behavior at low $T$ and short $q$ is the inconsistency of the Gaussian property. Thus, merely by probing the behavior of $\tau_{self}(q)$ at low $T$ and long times, we can conclude that the distribution of displacements is non-Gaussian, a result that we already proved by directly observing the exponential tail in the Self van Hove function. This \textit{additional} confirmation for the non-Gaussian statistics is not surprising since $F_{self}(q,t)$ essentially contains (via Fourier transform) the same information on the distribution of displacements as the Self van Hove function does~\cite{hansen,nagele}. Therefore, if motion is diffusive a behavior of $\tau_{self}(q)$ not compatible with $ \sim q^{-2}$ (in this case $\sim q^{-1}$) represents an alternative testimony on the non-Gaussian statistics of the distribution of displacements.\\

In order to gain intuition on the microscopic scenario that produces the $q^{-1}$ behavior of $\tau_{self}(q)$ at low $T$ and short $q$, we should notice that the $q^{-1}$ trend starts at those $q$ values at which we already documented the crossing between $\tau_{self}(q)$ and $\tau_{coll}(q)$ (\textit{i.e.}, around $q\sigma \lesssim 2$ for $T = 0.1025$, Fig.~\ref{Fig_tau_self_coll}).  This dynamic length scale was previously interpreted in terms of a linear size of cooperative domains of particles. Again, we can exploit the idea that the system at low $T$ is composed by a set of dynamic CR's. Taking into account the different mobility of the particles according to their number of bonds, we can speculate that the cores of the CR's are presumably formed by the less mobile (tightly bounded) particles whereas their borders would be delimited by the more mobile (weakly bounded) particles (this idea will be deeply explored in the next section). Thus, the CR's would continuously evolve by losing and gaining weakly bounded particles at the border, therefore changing the position of their borders. The characteristic time to change position (and also refresh the composition of the CR's with new particles) would be given by the crossing between $\tau_{self}(q)$ and $\tau_{coll}(q)$. We note on passing that the $q^{-1}$ behavior should not be associated with an individual particle ballistic motion, since the particle MSD for this regime is not quadratic but almost linear in time (see the MSD section). Nevertheless, we stress that, since we have a viscoelastic liquid, apart from the border displacement, the CR's will also be connected between each other at any time to transmit the network elasticity and therefore move in a coherent manner.\\

We should highlight that, to the best of our knowledge, this  $q^{-1}$ behavior has not been previously documented in such a clear way for a viscous system in equilibrium. Indeed, similar behaviors have only been documented and rationalized in out-of-equilibrium gelling systems as a result of their aging dynamics~\cite{cipelletti_2000,bouchaud_pitard,cipelletti_2005}. Finally, to close our discussion on the  $q$-dependence of $\tau_{self}(q)$, we should note that our heuristic scenario based on the emergence of a collective correlated dynamics should be systematically studied to not only reach a formal description but to also probe its $q$ extension to smaller wavevectors and for even lower $T$'s than those reached in the present study. In this respect, we should consider, as already anticipated in our discussion on the Self van Hove function, that as soon as the Gaussian behavior will be recovered (\textit{i.e.}, for times able to relax the equilibrium fluctuations at smaller $q$ values than those reached in the present simulation) the Central Limit Theorem will hold again, converging again to the diffusive regime ($q^{-2}D^{-1}$). Indeed, this seems to be the tendency for the smallest $q$ investigated here, for which $\tau_{self}(q)$ and $q^{-2}D^{-1}$ have almost reached a common value (see Fig.~\ref{Fig_tau_pannel}a) for $T = 0.1025$).\\

As done for $\tau_{self}(q)$, we now present separately the $q$-dependence of $\tau_{coll}(q)$ in Fig.~\ref{Fig_tau_pannel}b). As mentioned before, at any $T$ and for intermediate $q$, $\tau_{coll}(q)$ shows the typical oscillations in phase with the corresponding structure factor. Indeed the expected $\gamma q^{-2}S(q)$ behavior for the high $T$ liquid~\cite{pusey_1975,deGennes_1959} already discussed in  Fig.~\ref{Fig_tau_self_coll} seems to be a good qualitative approximation to describe the intermediate and large $q$-dependence of $\tau_{coll}(q)$. In particular at high $T$ (see $T = 0.14$ in Fig.~\ref{Fig_tau_pannel}b)) we can reproduce in a rather good qualitative manner the total $q$-dependence by taking $\gamma \cong D^{-1}$~\cite{nagele}. The most intriguing behavior arrives at low $T$ and short $q$. At low $T$ and around $q\sigma \in [2,3]$, $\tau_{coll}(q)$ starts to develop a shoulder which is placed again at the crossing between  $\tau_{coll}(q)$ and $\tau_{self}(q)$ (linear size of the CR's). We should notice that this shoulder is indeed a dynamic signature which is not contained in the Structure factor (see Fig.~\ref{Fig_Sq_total}). At even smaller wavevectors (where $\tau_{self}(q) \sim q^{-1}$, $q\sigma \lesssim 1.5$),  $\tau_{coll}(q)$ significantly drops reaching an almost constant value. However, the precise mechanism governing the constant relaxation time of the collective microscopic dynamics at such as short wavevectors is still unclear.\\
   
\subsection{Cooperative Domains: A Structural Signature}

We have explored the dynamics of the system by paying special attention to its relaxation time at different spatial scales. In this respect, we have found a distinctive length scale which characterizes the collective dynamics  and increases upon cooling the system. This length scale has been interpreted as a linear CR size. So far, these CR's would conform the microscopic picture of the low $T$ liquid from a purely dynamic viewpoint. However, it is not clear whether or not the dynamic CR picture also manifests as a structural property.\\

\begin{figure}[tb]
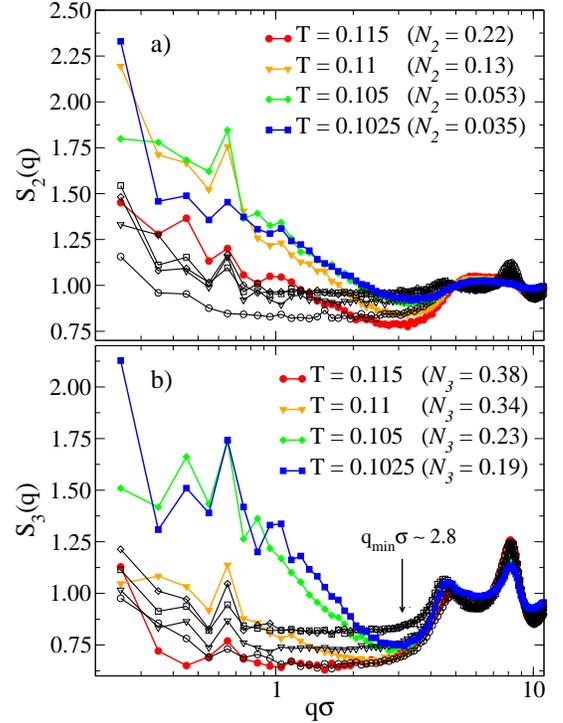

\includegraphics[clip,width=0.82\linewidth]{Sq_n2.eps}
\includegraphics[clip,width=0.82\linewidth]{Sq_n3.eps}
\caption{Partial structure factor, $S_n(q)$, for $n\in{\lbrace2,3\rbrace}$ at different $T$'s: a) Log-linear plot of the structure factor, $S_2(q)$, of those particles having two bonds and b)  Log-linear plot of the structure factor, $S_3(q)$, of those particles having three bonds. In b) we also marked with a vertical arrow the position of the local minimum at which a long range correlation start for $T = 0.1025$ ($q\sigma \cong 2.8$).  Both figures also show for each $T$ the structure factor of a set of randomly chosen particles whose number of particles coincides with the number of particles of the corresponding partial structure factor at the same $T$ (black lines with symbols as in the corresponding partial structure factor). Included in the legend of both figures are the fraction of particles for each $T$ (${\cal N}_n$ ; $n\in{\lbrace2,3\rbrace}$).}
\label{Fig_Sq_n}
\end{figure}
    
To further explore the possibility of a long range \textit{static} spatial correlation we proceed by assuming that the correlated distribution of particles with $n$ bonds previously discussed has a static counterpart, originating regions formed by completely bounded particles, surrounded by particles with one or more broken bonds. This idea (already anticipated in the previous section) would also be coherent with our interpretation of the system connectivity (section III.A), where we already interpreted the system bonding dependence at low $T$ in terms of a spatial localization of particles according to their number of bonds. We should also notice that this picture is indeed reminiscent of the structural signatures expected in the vicinity of a liquid-liquid transition, a possibility first proposed for liquid water~\cite{poole1992phase}, and later for supercooled silicon~\cite{sastry_silicon}, which has been recently demonstrated to be a generic feature of tetrahedral networks~\cite{water1_ff}.\\
     
Motivated by these considerations we have separated the total structure factor of the system, $S(q)$, into partial structure factors of particles with a given number of bonds, $S_n(q)$. In this respect, Fig.~\ref{Fig_Sq_n}a) and b) show  $S_n(q)$ at different $T$'s for those populations constituted by particles with $n=2$ and $n=3$ bonds respectively. Certainly, and according to the idea exposed in the previous paragraph, we could also expect to detect a structural pattern for the $n=0$ and $n=1$ populations. However, at low $T$ these populations are too small to give us a reliable statistics (${\cal N}_n < 0.01$ for $n\in{\lbrace0,1\rbrace}$ at $T = 0.1025$). In addition, we could also consider
the population of four-bounded particles, however they represent the majority of the particles in the system at low $T$ (${\cal N}_4 = 0.76$ and ${\cal N}_4 = 0.70$ for $T = 0.1025$ and $T = 0.105$ respectively). Thus, we also omit $S_4(q)$ since its long range structure is partially hidden by the underlying total structure given by the total $S(q)$. To identify the structure of the different $S_n(q)$ as a distinctive structural property of a given population, we have also included in Fig.~\ref{Fig_Sq_n}a) and b) the structure factors of different sets of randomly chosen particles whose number of particles coincides with the number of particles of the corresponding partial structure factor, $S_n(q)$, at the same $T$. For instance, for $T = 0.1025$ we have ${\cal N}_2 = 0.035$ (Fig.~\ref{Fig_Sq_n}a)) and ${\cal N}_3 = 0.19$ (Fig.~\ref{Fig_Sq_n}b)), therefore we added in the corresponding figure the structure factor of a set of randomly chosen particles with $0.035 N$ and $0.19 N$ particles respectively as obtained from the same simulation at $T = 0.1025$ (black lines with empty squares in both Fig.~\ref{Fig_Sq_n}a) and b)).\\

In Fig.~\ref{Fig_Sq_n}a) and for intermediate $q$ we do not clearly see the most obvious structure (main and tetrahedral peaks) that we have in the total $S(q)$ at the same $T$'s (Fig.~\ref{Fig_Sq_total}). This is obviously due to the small number of particles forming the $n=2$ population (e.g. ${\cal N}_2= 0.053$ at $T=0.105$ and ${\cal N}_2= 0.035$ at $T=0.1025$). However at short $q$ (typically $q\sigma \lesssim 3$) $S_2(q)$ starts to increase upon cooling the system. Despite some noise at very short $q$, this small-$q$ structure is characteristic of the $n=2$ population since the corresponding randomly chosen sets of particles show a structure factor which does not reveal any long range structure, being almost crowded around $S(q) = 1$. This effect is more pronounced for the $n=3$ population, Fig.~\ref{Fig_Sq_n}b), for which ${\cal N}_3$ has a statistically reliable value for all $T$'s represented. In this case, and due to the significant value of ${\cal N}_3$, $S_3(q)$ shows part of the intermediate $q$ structure present in the total $S(q)$ (tetrahedral and main peaks). Again, we clearly see how upon cooling the system $S_3(q)$ reveals at short $q$ and low $T$ a clear long range structure: $S_3(q)$ monotonically increases as $q \rightarrow 0$. This behavior is not present in the corresponding randomly chosen sets of particles, which again remain crowded around $S(q) = 1$.\\

\begin{figure}[tb]
\center
\includegraphics[width=0.95\linewidth]{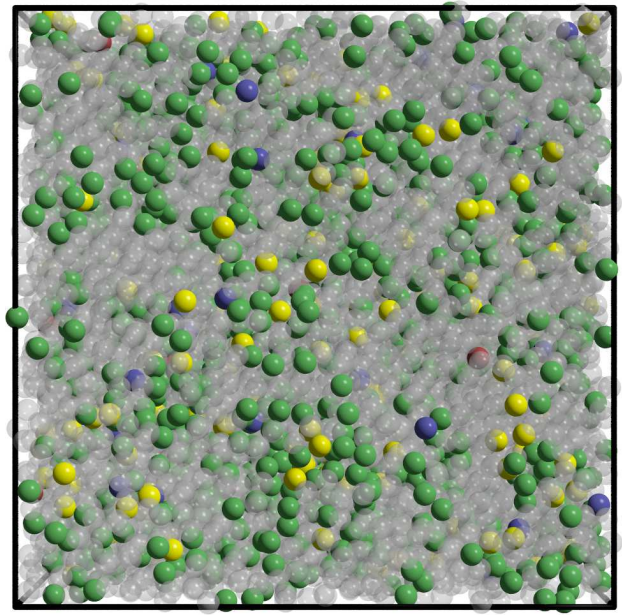}
\caption{Configuration snapshot at $T = 0.1025$. The code for the number of bonds-color in the figure is:  $n = 4$ - grey, $n = 3$ - green, $n = 2$ - yellow, $n = 1$ - blue, and $n = 0$ - red. The fully-bounded particles are partially transparent for the sake of clarity.}
\center
\label{Fig_cluster_snapshot}
\end{figure}

At this point, it is tempting to link the structural pattern observed at short $q$ and low $T$ for the different $S_n(q)$ represented in  Fig.~\ref{Fig_Sq_n}a) and b) to the dynamic picture previously discussed. Thus, in particular, we see how $S_3(q)$ at $T = 0.1025$ starts to increases upon decreasing $q$ from a local minimum placed around $q\sigma \cong 2.8$ (see Fig.~\ref{Fig_Sq_n}b)). This value almost coincides with the dynamic length scale, $\xi(T) \sim q^{*}(T)^{-1}$, detected by the $\tau_{self}(q)$-$\tau_{coll}(q)$ crossing and by the shoulder in $\tau_{coll}(q)$ at the same $T$ (see Fig.~\ref{Fig_tau_self_coll} and Fig.~\ref{Fig_tau_pannel}b)), \textit{i.e.} the dynamic length scale that we associated to the CR's. With this idea in mind, the short $q$ behavior at low $T$ of $S_2(q)$ and $S_3(q)$ would indicate the long range inter-domain structure, which indeed would start at the typical \textit{static} domain size ($q\sigma \lesssim 3$ at low $T$).\\ 

By the procedure presented here we can indeed establish a value for a static length at low temperature (typically $T = 0.1025$ and $T = 0.105$) which is quantitatively similar to that obtained for the dynamic length scale at the same $T$. However, the current data do not allow us to obtain accurate values for the static length at higher temperatures. For that reason we can merely compare the dynamic and static length scales at low $T$ but we cannot establish whether or not this correlation is maintained at higher $T$'s. Despite the physical consistence of the equivalence between dynamic and static lengths present in the system at low $T$ needs to be rationalized, it is clear that with the methodology exposed here we can document the existence of both lengths, with values quantitatively compatible at low $T$. Finally we point out that the short $q$ behavior of $S_2(q)$ and $S_3(q)$ at low $T$ should still be explored in a more systematic way to better characterize its functional behavior, which here seems to be compatible with a power law behavior, $S(q) \sim q^{- \alpha}$ with $\alpha > 0$, characteristic of a fractal system~\cite{martin_hurd,schaefer_fractal,carpineti_fractal}.\\ 

Finally, to illustrate the static correlations, we present in  Fig.~\ref{Fig_cluster_snapshot} a snapshot corresponding to a given configuration of the system at $T = 0.1025$. The figure clearly shows that fully-bounded particles (depicted in transparent grey) tend to cluster into core domains which are surrounded by less-bounded particles (green, yellow, blue, and red particles), which indeed would form the border of the domains. The snapshot thus provides a visual support to the presence of the static CR's of fully-connected particles.

\section{SUMMARY AND CONCLUSIONS}

We have presented a study on the dynamics and structure of a system of tetravalent patchy particles by means of Brownian Dynamics simulations~\cite{John_valence,rovigatti_molphys}. We have explored in depth the equilibrium temperature evolution of the system under different static and dynamic views with the final scope of proposing a microscopic picture for the viscoelastic nature of the low-temperature liquid. In this respect, we have exploited the highly directional interaction between the particles for describing the system connectivity in a very precise manner where the particles have been discriminated into different populations according to their number of bonds. This discrimination has allowed us to perform a rich description of the system by which we have revealed a clear link between connectivity, particle mobility, and structure.\\

We have shown how in the system the distribution of the potential energy per particle can be interpreted in terms of a penta-modal distribution of bonds per particle. By this distribution of bonds we have thoroughly characterized the increasing connectivity of the system upon decreasing temperature. In particular, we have proved how at high $T$ the distribution of bonds can be understood in terms of a binomial distribution which is based on the assumption of an independent bonding process. However, we have also seen how, upon cooling the system, the assumption of bond independence breaks down. Thus, at low $T$, it appears a bond correlation that entropically favors the presence of more weakly bounded particles than those expected from an independent bond formation process.\\

We have also performed an extensive study on the temperature evolution of the individual and collective dynamics exploring its relation with the system connectivity. For instance, we have discussed the low-$T$ Arrhenius evolution of the diffusion coefficient, $D$. In this respect, we have introduced a novel methodology to discriminate the mean square displacement into different populations of particles according to their number of bonds at $t = 0$. By this methodology we have established a clear relation between number of bonds and particle mobility. Thus, we have followed the partial mean square displacement of the different populations of particles for different temperatures and over a long time window where different regimes (\textit{e.g.} super and sub-diffusive) have been documented for the different populations.\\ 

We have carried out a similar analysis for the complete distribution of displacements by intensively exploring the van Hove function of the system. At high and intermediate temperatures, the partial and total Self van Hove functions show the Gaussian behavior expected for a high-$T$ liquid. However, at low $T$ we have found a noticeable non-Gaussian behavior manifested through a long exponential tail which has been characterized for the different populations of particles at different time scales. In particular, our methodology has allowed us to report the intermittent dynamics of the weakly bounded particles at short times, which manifests itself by large displacements (jumps) at the single particle level. Despite at low $T$ and long times the dynamics is already diffusive, the non-Gaussian behavior of the Self van Hove function has permitted to document in a very clear way the \textit{anomalous yet Brownian diffusion} present in the system~\cite{pnas_non_gaussian_granick,nat_mat_non_gaussian_granick,hs_non_gaussian}. Our study has been complemented with an alternative procedure,  based on the non-Gaussian parameter, to quantify the non-Gaussian statistics. In addition, we have analyzed the collective dynamics of the system in real space by means of the Distinct van Hove function. In this respect, we have shown the notorious rigidity of the low-$T$ liquid, which appears significant when compared with other strong and fragile glass forming-liquids~\cite{walter_andersen,sciortino_sup_water,horbach_walter_silica,horbach_walter_silica_2}.\\

We have also investigated the dynamics of the system in Fourier space through the Self and Collective scattering functions. In this respect, we have covered several temperatures within a large range of the wavevector, $q$, by analyzing the $q$-dependence of the self and collective relaxation times, $\tau_{self}(q)$ and $\tau_{coll}(q)$. In particular, at low $T$ we have shown how $\tau_{coll}(q)$ and $\tau_{self}(q)$ decouple at intermediate spatial scales ($q\sigma \gtrsim 3$), where $\tau_{coll}(q)$ exceeds $\tau_{self}(q)$ by more than one order of magnitude. Thus, the system retains its dynamic collective correlations at spatial scales even larger than that associated to the tetrahedral distance, despite the particles having individually moved even larger distances. At short $q$, $\tau_{self}(q)$ finally overpasses $\tau_{coll}(q)$ at a certain $q$ value which we have identified as a dynamic length scale which increases upon cooling the system. The dynamic length scale is linked to the size of dynamic domains of particles which evolve in a cooperative manner, and reaches a value of the order of few particle diameters at the lowest $T$ investigated.\\

Looking into the $q$-dependence of $\tau_{self}(q)$, we have shown how at high temperatures $\tau_{self}(q)$ reproduces the behavior expected for a high-$T$ liquid, \textit{i.e.} it shows a $q^{-2}D^{-1}$ behavior within the explored $q$ range. However, at low $T$ we have documented the emergence of a $q$-independent behavior at intermediate $q$ and a $q^{-1}$-dependence at short $q$ ($q\sigma \lesssim 2$). In particular, the  $q^{-1}$-dependence of $\tau_{self}(q)$ appears as an interesting feature of the low $T$ dynamics of the system which so far had only been detected in out-of-equilibrium gelling systems as a consequence of their aging dynamics~\cite{cipelletti_2000,bouchaud_pitard,cipelletti_2005}.\\

Finally, we have investigated the emergence of long-range static correlations compatible with the detected dynamic length scale. We again took advantage of our methodology to separate the low $T$ structure factor of the system into partial structure factors according to the particle number of bonds. In order to reveal the cooperative-domain picture suggested by our dynamic results, we have studied the structure factor of those populations of particles with $n = 2$ and $n =3 $ bonds to reveal the \textit{static} extension of the domains. Indeed, at low $T$ we have found that a structure is present at small $q$, where the partial structure factors show a monotonically increasing behavior as $q \rightarrow 0$. Interestingly, the values at which the small-$q$ structure emerges are very similar to those reported for the long range dynamic length scale ($q\sigma \lesssim 3$), suggesting a plausible connection between structure and dynamics which extends to large spatial scales.\\

In summary, we have documented the rich dynamic and structural phenomenology present in a tetrahedral network liquid of tetravalent patchy particles. The resulting microscopic scenario has been interpreted in terms of a dynamic-static domain picture which accounts for the viscoelastic nature of the low $T$ liquid. This picture relies on the basis of a viscous flow of cooperative domains of particles where the network elasticity is mediated by the inter-domain connections. To reach our description of the low-$T$ liquid we have supported our investigation with a new methodology consisting on the discrimination of the system connectivity in terms of  particle populations, where each population corresponds to all the particles with a given number of bonds. We believe that the methodology employed in this work could reveal similar results and even new insights in other gel- and glass-forming liquids.\\
 
\section*{ACKNOWLEDGEMENTS}
L.R. acknowledges support from the Austrian Research Fund (FWF) through the Lise-Meitner Fellowship M1650-N27 and from the European Commission
through the Marie Sk\l{}odowska-Curie Fellowship 702298-DELTAS. F.S. acknowledges support from ETN-COLLDENSE (H2020-MCSA-ITN-2014, Grant No. 642774).  We thank Walter Kob for many illuminating discussions.

\appendix*
\renewcommand\thefigure{A.\arabic{figure}}
\setcounter{figure}{0}

\section{STRUCTURE AND DYNAMICS IN FOURIER SPACE}
\label{Appendix}

In this Appendix we present results on the static structure factor of the system as well as some crude data corresponding to the Self and Collective scattering functions for supporting part of the discussion present in the main text. In addition we present the protocol for obtaining the relaxation times $\tau_{self}(q)$ and $\tau_{coll}(q)$ and discuss their reliability at low $T$ and short $q$.\\ 

\begin{figure}[tb]
\center
\includegraphics[width=0.77\linewidth]{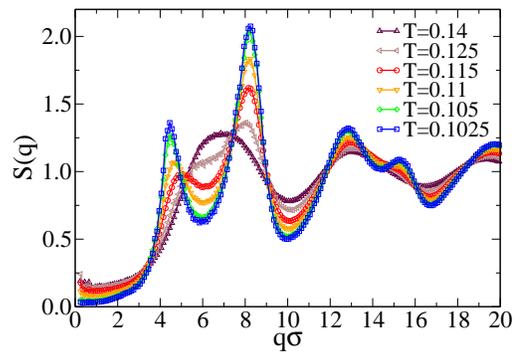} 
\caption{Structure Factor, $S(q)$, for the total number of particles at different temperatures.} 
\center
\label{Fig_Sq_total} 
\end{figure}

Fig.~\ref{Fig_Sq_total} shows the structure factor, $S(q)$, of the system at different temperatures (a previous discussion on the structure factor of this system has been already presented in Ref.~\cite{rovigatti_molphys}). Whereas at high temperatures ($T = 0.14$) the structure of the system is essentially that corresponding to a hard-sphere liquid with a main peak at $q\sigma \cong 7$, upon cooling the system we clearly see the signature of an archetypal tetrahedral structure by the emergence of an additional tetrahedral peak at $q\sigma \cong 4.5$.  Thus, due to the judicious choice of the density and patch size, the system at low $T$ exhibits no crystalline order but an amorphous tetrahedral network structurally similar to that present in atomistic systems such as silica or silicon~\cite{binder_walter,horbach_walter_silica}, classical water models ~\cite{ivan_patchy_water}, or models of colloidal gels where double peak structures are imposed by means of three-body interactions~\cite{gado_walter_gel}.\\

\begin{figure}[tb]
\center
\includegraphics[clip,width=0.8\linewidth]{f_self_T0.1025.eps}\vspace{0.2cm}
\includegraphics[clip,width=0.78\linewidth]{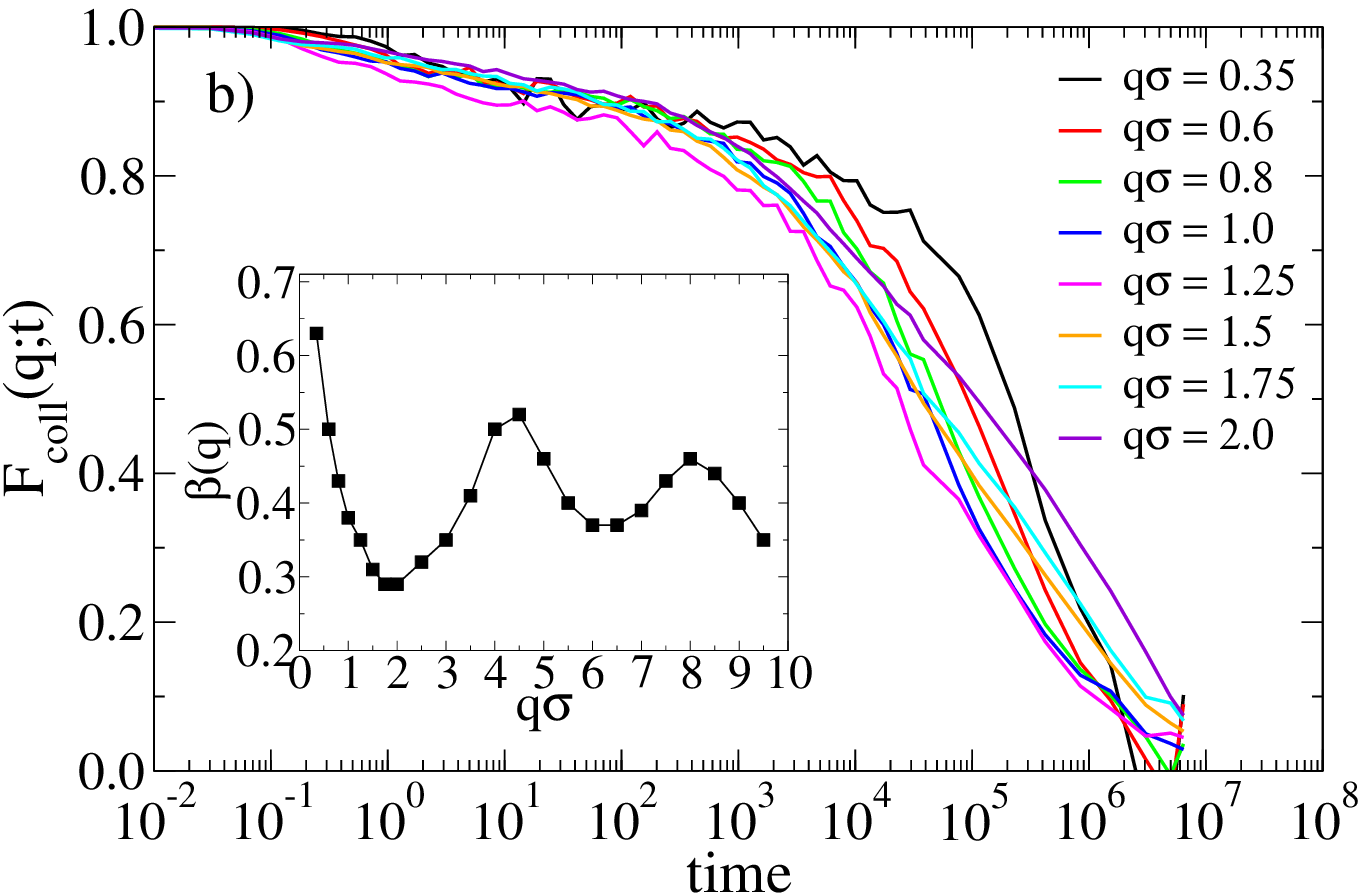}
\caption{a) Linear-Linear plot of log$(F_{self}(q,t))$ for $q\sigma \in [0.35,2]$ at $T=0.1025$ (we added solid symbols to highlight the longest correlation times), b) $F_{coll}(q,t)$ for $q\sigma \in [0.35,2]$ at $T=0.1025$. Inset in b): $\beta(q)$ exponents of the stretched exponential fit (Equation (A.1)) as a function of $q$.}
\center
\label{Fig_f_self_coll_T0.1025}
\end{figure}

Concerning the collective and individual relaxation of the system, we measure the corresponding relaxation times associated to the Self and Collective scattering functions, $\tau_{self}(q)$ and $\tau_{coll}(q)$, by fitting the final $\alpha$-decay of the corresponding correlation function by means of a stretched exponential function~\cite{hansen} of the form $e^{-(\tau/\tau_\alpha(q))^{\beta(q)}}$ and then integrate over the time domain:

\begin{small}
\begin{eqnarray}
\tau_{dyn}(q) = \int_0^\infty e^{-(\tau/\tau_\alpha(q))^{\beta(q)}} dt=\frac{\tau_\alpha(q)}{\beta(q)}\Gamma\left(\frac{1}{\beta(q)}\right)
\end{eqnarray}
\end{small}
\\
\noindent
Where $dyn\in{\lbrace self,coll \rbrace}$ whereas $\tau_\alpha(q)$ and $\beta(q)$ are the two parameters entering into the stretched exponential fit for a given $q$ value, $\Gamma(x)$ being Euler's Gamma function.\\

In order to show the reliability of our results for $\tau_{self}(q)$ and $\tau_{coll}(q)$ at low $T$ and short $q$, we show  in Fig.~\ref{Fig_f_self_coll_T0.1025} the correlation functions from which we extracted $\tau_{self}(q)$ and $\tau_{coll}(q)$ for the lowest $T$ and the shortest $q$'s investigated. In particular Fig.~\ref{Fig_f_self_coll_T0.1025}a) shows log$(F_{self}(q,t))$ in a double linear plot. Despite at short $q$ our correlators do not completely decay (e.g. $F_{self}(q\sigma = 2;t_{long}) \cong 0.2$), their time dependence (typically $q\sigma \lesssim 1$) seems to be consistent with a pure exponential decay (see the linear time dependence of log$(F_{self}(q,t))$ for $q\sigma \lesssim 1$), that is, they already seem to present the Debye decay expected for the $q \rightarrow 0$ limit. In addition,  Fig.~\ref{Fig_f_self_coll_T0.1025}b) shows $F_{coll}(q,t)$ at $T = 0.1025$ for the shortest explored $q$ values. We see how the different $F_{coll}(q,t)$ are almost crowded, indicating that their $\tau_{coll}(q)$ values are quite similar ($q$-independent regime for $q\sigma \lesssim 1.5$, blue curve in Fig.~\ref{Fig_tau_pannel}b)). We can also notice that from $q\sigma \geq 1.5$, $F_{coll}(q,t)$ starts to relax more slowly in consistence with the emergence of the shoulder documented in Fig.~\ref{Fig_tau_pannel}b). We also present in the inset of Fig.~\ref{Fig_f_self_coll_T0.1025}b) the corresponding $\beta(q)$ exponents as obtained from our stretched exponential fit (Equation (A.1)). At intermediate $q$, $\beta(q)$ shows the typical oscillations in phase with the structure factor. In addition, $\beta(q)$ starts to increase upon decreasing $q$ after reaching a minimum placed at $q\sigma \cong 2$, a value which is again close to the $\tau_{self}(q)$-$\tau_{coll}(q)$ crossing, and compatible with the linear size of the correlated regions, (Fig.~\ref{Fig_tau_self_coll}), pointing that for this value the system presents its maximum degree of heterogeneity. From this maximum $\beta(q) \rightarrow 1$ as $q \rightarrow 0$, indicating that the degree of heterogeneity tends to disappear at large spatial scales.

\bibliographystyle{apsrev}

\end{document}